\documentclass[preprint,showpacs,preprintnumbers,amsmath,amssymb]{revtex4}
\usepackage{graphicx}
\graphicspath{{./Figs/}}
\usepackage{epsfig}
\usepackage{bm}
\usepackage{amsfonts}
\usepackage{subfigure}
\usepackage{multirow}
\usepackage{float}
\usepackage{color}
\usepackage{xcolor, soul}

\providecommand{\tabularnewline}{\\}

\newcommand{\e}{\epsilon}

\newcommand{\be}{\begin{equation}}
\newcommand{\ee}{\end{equation}}
\newcommand{\ba}{\begin{eqnarray}}
\newcommand{\ea}{\end{eqnarray}}

\newcommand{\rd}{{\rm d}}
\newcommand{\Mpl}{M_{\rm Pl}}

\usepackage{amsmath}

\usepackage{graphicx,color}

\usepackage{dcolumn}
\usepackage{hyperref}
\allowdisplaybreaks[1]

\sethlcolor{green}
\providecommand{\tabularnewline}{\\}	

\begin{document}
	
	\title{\textbf{Primordial black holes in SB SUSY Gauss-Bonnet inflation}}
	
%ashrafzadeh.ali@gmail.com
	
	\author{A. Ashrafzadeh\footnote{a.ashrafzadeh@uok.ac.ir}, M. Solbi\footnote{miladsolbi@gmail.com}, S. Heydari\footnote{s.heydari@uok.ac.ir} and K. Karami\footnote{kkarami@uok.ac.ir}}
	\affiliation{\small{Department of Physics, University of Kurdistan, Pasdaran Street, P.O. Box 66177-15175, Sanandaj, Iran}}
	
	\date{\today}
\begin{abstract}
	
Here, we explore the formation of primordial black holes (PBHs) within a scalar field inflationary model coupled to the Gauss-Bonnet (GB) term, incorporating the low-scale spontaneously broken supersymmetric (SB SUSY) potential.
The coupling function amplifies the curvature perturbations, consequently leading to the formation of PBHs and detectable secondary gravitational waves (GWs).
Through the adjustment of the model parameters, the inflaton can be decelerated  during an ultra-slow-roll (USR) phase, thereby augmenting curvature perturbations. Beside the observational constraints, the swampland criteria are investigated.
Our computations forecast the formation of PBHs with masses around ${\cal O}(10)M_{\odot}$, aligning with the observational data of LIGO-Virgo, and PBHs with masses ${\cal O}(10^{-6})M_{\odot}$ as potential explanation for the ultrashort-timescale microlensing events recorded in the OGLE data.
Additionally, our proposed mechanism can generate PBHs with masses around ${\cal O}(10^{-13})M_{\odot}$, constituting roughly 99$\%$ of the dark matter.
The density parameters of the produced GWs ($\Omega_{\rm GW_0}$) intersect with the sensitivity curves of GW detectors. Two cases of our model fall within the nano-Hz frequency regime. One of them satisfies the power-law scaling as $\Omega_{\rm{GW}}(f) \sim f^{5-\gamma}$, with the $\gamma = 3.51$, which is consistent with the data of NANOGrav 15-year.

\end{abstract}
	
	\pacs{ }
	
	\maketitle
	
	%---------------------------------------------------------------------------------------------
	\newpage
	%%%%%%%%%%%%%%%%%%%%%%%%%%%
	%%%%%%%%%%%%%%%%%%%%%%%%%%%
	\section{Introduction}
	\label{sec1}
	%%%%%%%%%%%%%%%%%%%%%%%%%%%
	%%%%%%%%%%%%%%%%%%%%%%%%%%%
%
Zel'dovich and Novikov for the first time proposed the formation of PBHs arising from the enhanced curvature perturbations \cite{zeldovich:1967}.
The subsequent researches have shown that PBHs with masses less than $10^{15} g$ would have evaporated, while more massive PBHs might still exist today \cite{Hawking:1974,Hawking:1976,Hawking:1971,Page:1976,carr:1975,carr:1974}.
The generated PBHs in the early universe, could potentially be viable candidates for the dark matter
~\cite{Olmo:2011,Cruz-Dombriz:2012,Chapline:1975,Ivanov:1994,Polnarev:1985,Qiu:2023NEWQiu,Teimoori-b:2021,Solbi-b:2021,Rezazadeh:2021,Solbi-a:2021,YCai:2021,YCai:2023,AChakraborty:2022,GDomenech:2023,TPapanikolaou:2022,TPapanikolaou:2023a,TPapanikolaou:2023b,SChoudhury:2014}.
The successful detection of GWs from merging black holes, by LIGO-Virgo teamwork has significantly increased interests in PBHs \cite{Abbott:2016,Abbott:2016-a,Abbott:2016-b,Abbott:2017-a,Abbott:2017-b,Abbott:2017-c,Abbott:2017,Abbott:2019}.
The source of the GWs detected by LIGO-Virgo might be PBHs with mass scale of ${\cal O}(10)M_{\odot}$. Moreover, the ultrashort-timescale microlensing events observed in the OGLE data may be explained by PBHs with masses close to ${\cal O}(10^{-5})M_{\odot}$ \cite{OGLE}.
Also, PBHs within the mass range ${\cal O}(10^{-16}-10^{-11})M_{\odot}$, could make up the entirety of the dark matter content~\cite{Gould:1992,Alcock:2001,Teimoori:2021,Heydari:2021,Heydari:2024,Heydari:2022,Heydari:2023a,Heydari:2023b,Solbi:2024,Dalcanton:1994,Laha:2020,Ali-Haimoud:2017,Sato-Polito:2019,Wang:2018,Jacobs:2015}.

PBH formation needs a significant amplitude of the primordial curvature perturbations to generate overdense zones in the radiation-dominated (RD) phase. Upon horizon re-entry of the boosted scales, the overdense zones could gravitationally collapse to produce PBHs.
Hence,
the formation of PBHs requires an increase in the scalar power spectrum to the order  $\mathcal{O}(10^{-2})$ on small scales.
The amplitude of the scalar power spectrum at the pivot scale $k_{*}=0.05~ \mathrm{Mpc}^{-1}$ is defined from the Planck measurements as  $2.1 \times 10^{-9}$ \cite{akrami:2020}.
Consequently, an amplification to seven orders of magnitude in the scalar power spectrum, on small scale, is required to produce PBHs \cite{Mishra:2020,Fu:2019,Kawai:2021,Dalianis:2019,mahbub:2020,Kawaguchia:2023,Villanueva-Domingo:2021,Ashrafzadeh:2024,Gao:2021,Lin:2020,Kamenshchik:2019,Zhang:2022,SChoudhury:2023}.

As amplified scales re-enter the horizon during the RD epoch, concurrently with PBH formation, the secondary GWs could be propagated.
Therefore, detection of the signals of these GWs offers a promising technique for indirect identification of PBHs \cite{Bhattacharya:2023,GDomenech:2021a,GDomenech:2021b,GDomenech:2024,SMittal:2022,MRGangopadhyay:2023b,GBhattacharya:2023,Banerjee:2022NEWPapanikolaou,Papanikolaou:2024NEWPapanikolaou}.
In other words, the collapse of amplified curvature perturbations upon the horizon re-entry generates significant metric perturbations. At the second order, the scalar and tensor perturbations can interact, unlike the linear level. As a result, the second-order scalar metric perturbations can produce  the GWs background~\cite{Basilakos:2024NEWPapanikolaou,Tzerefos:2023NEWPapanikolaou,Braglia:2021NEWBraglia,Braglia:2022NEWBraglia,Braglia:2024NEWBraglia,BartoloPRL:2019,BartoloPRD:2019,CaiPRL:2019,Fumagalli:2021,CaiJCAP:2019,Wang:2019,CaiPRD:2019}.
Since the NANOGrav signal remains unconfirmed, it is difficult to propose plausible mechanisms for its origin. In this regard, the NANOGrav pulsar timing array team has recently unveiled the background GWs within 15 years data~\cite{Agazie1:2023,Agazie2:2023,Agazie3:2023,Agazie4:2023}.

One widely utilized strategy for amplifying the scalar power spectrum involves  a brief USR phase.
The inflaton undergoes an effective deceleration during the USR phase,
consequently resulting in the substantial increase in the curvature perturbations \cite{Ragavendra:2023}.
Hence, the USR era is essential for PBH formation and can be accomplished by numerous means.
Under the slow-roll conditions, the scalar power spectrum does not increase enough to produce PBHs. Conversely, during the USR phase, a decrease in the inflaton velocity provides the sufficient time for amplification of the primordial curvature perturbations, which can subsequently result in the PBH formation after re-entering the horizon.
This process can be realized within the framework of both the standard inflationary models and the modified gravity theories.
In the standard model of inflation, an inflection point in the potential can lead to the significant deceleration of the inflaton  and consequently, generation of the USR phase \cite{Mishra:2020,Dalianis:2019,mahbub:2020,Dimopoulos:2017}.
Furthermore, the requirements for attaining the USR epoch within the modified gravity framework can be achieved by selecting a suitable coupling function and optimizing the model parameters \cite{Fu:2019,Kawai:2021,Kawaguchia:2023,Villanueva-Domingo:2021,Gao:2021,Lin:2020,Kamenshchik:2019,Zhang:2022,SChoudhury:2023}.
In this regard, it has been demonstrated that PBHs can be generated within the frameworks of non-minimal coupling and non-canonical models through the implementation of proper coupling functions
\cite{Fu:2019,Gao:2021,Lin:2020,Kamenshchik:2019}.

In recent years, numerous studies have focused on the higher dimensional gravity to understand the low-energy limit of the string theory. The examination of the GB theory could offer a comprehensive framework for addressing the conceptual issues related to gravity. The GB gravity, originally proposed by Lanczos~\cite{Lanczos:1938} and subsequently proclaimed by David Lovelock~\cite{Lovelock:1971}, represents a significant higher-dimensional generalization of the Einstein gravity~\cite{Lanczos:1938,Lovelock:1971,RashidiNozari:2020,AziziNozari:2022,ShahraeiniNozari:2022}. In the GB theory, the order of the field equations does not exceed the second order, and the theory is free of ghost. Recently, numerous attempts have been made to derive black hole solutions. However, the first exact black hole solution in the GB gravity was calculated by Boulware and Deser~\cite{Boulware:1985,Myers:1988}. Subsequently, other researchers have explored exact black hole solutions with their thermodynamic properties~\cite{Cho:2002,Sahabandu:2006}. Additionally, various black hole solutions with matter sources, extending the Boulware-Deser solution, have been discovered in~\cite{HabibMazharimousavi:2009,Ghosh:2014}.
In recent years, researchers have investigated the formation of PBHs within single field inflationary model coupled with the GB term. Through these works, they have been able to revive the Higgs, natural, and fractional power-law potentials. The coupling functions in   modified gravity models with the GB term can enhance
the primordial scalar perturbations on small scales,
to produce
PBHs and GWs
\cite{Kawaguchia:2023,Kawai:2021,Ashrafzadeh:2024,Zhang:2022}.

In the next, the inflationary models are evaluated with swampland criteria. These criteria, derived from the string theory, include the distance conjecture and the de Sitter conjecture. The validity of the de Sitter conjecture is not consistent with the slow-roll conditions in the standard inflation. Nevertheless, in the modified inflationary models, both criteria can be  satisfied~\cite{Kehagias:2018,Garg:2019,Das:2019,Ooguri:2019}.

In this research, we aim to study the low-scale SB SUSY potential, which does not agree with the Planck observations in the standard model. Within the framework of modified gravity model with the GB term, we will examine the compatibility of the model with the latest observational constraints in addition to the swampland criteria thereto analyzing the PBHs production. Furthermore, we will investigate the GWs generated within the sensitivity range of the detectors. The structure of this paper is outlined as follows. Sect. \ref{sec2} presents a summary of the GB
Model. In Section \ref{sec3},
the method employed for enhancing the scalar perturbations on small scales is introduced.
In Section \ref{sec4}, the fractional abundances and various masses of PBHs will be computed. In Section \ref{sec5}, the energy spectra of GWs will be
estimated.
At the end, the summarized results are presented in Section \ref{sec6}.

%%========================section 2 =========================
%Inflationary
\section{Summary of Gauss-Bonnet Model}
\label{sec2}
We regard the Einstein-Gauss-Bonnet action in the following
\cite{Jiang:2013,Koh:2014,Guo:2010,Odintsov:2020,Gao:2020,AziziNozari:2022,ShahraeiniNozari:2022,RashidiNozari:2020,HAKhan:2022,MRGangopadhyay:2023a,SDOdintsov:2020,SDOdintsov:2023,SNojiri:2023a,EOPozdeeva:2020}
	\be
	{\cal S}=\int {\rm d}^4 x \sqrt{-g} \left[\frac{\Mpl^{2}}{2}R
	-\frac{1}{2}g^{\mu \nu}\nabla_{\mu}\phi
	\nabla_{\nu}\phi-V(\phi)-\frac{\xi(\phi)}{2}R_{\rm GB}^{2}\right],
	\label{action1}
	\ee
where $R$ is the Ricci scalar, $\Mpl=(8\pi G)^{-1/2}$ is the reduced Planck mass, $R^2_{\rm GB} \equiv R_{\mu\nu\rho\sigma}R^{\mu\nu\rho\sigma} - 4R_{\mu\nu}R^{\mu\nu} + R^2$ signifies the invariant scalar GB term and $\xi(\phi)$ denotes the coupling function between
the GB term and the scalar field $\phi$.
In this context, we examine the spatially flat Friedmann-Lema\^{\i}tre-Robertson-Walker (FLRW) metric.

Using  the derivation of the action (\ref{action1}) versus the metric $g^{\mu \nu}$ and scalar field $\phi$, one can derive the Friedmann equations and the equation of motion as follows
\ba
	\label{bge1}
	&& 3{\Mpl^{2}}H^2 = \frac12\dot{\phi}^2+V(\phi)+12\dot{\xi}H^3, \\
    \label{bge2}
    && -2{\Mpl^{2}}\dot{H} =\dot{\phi}^2 - 4\ddot{\xi}H^2 - 4\dot{\xi}H\big(2\dot{H} - H^2\big), \\
    \label{bge3}
    && \ddot{\phi}+3H\dot{\phi} + V_{,\phi}= - 12\xi_{,\phi}H^2\big(\dot{H}+H^2\big),
    \ea
in which, $H\equiv \dot{a}/a$ represents the Hubble parameter, $a$ is the scale factor, the dot  $({,\phi})$ indicates  derivation with respect to the cosmic time $t$ ( the scalar filed $\phi$).

In the GB model, the slow-roll parameters are given by
\cite{Jiang:2013,Koh:2014,Guo:2010,Odintsov:2020,Gao:2020}
	\ba
	\label{SRH}
    \e_1\equiv\frac{-\dot{H}}{H^2},~~~\e_2\equiv\frac{\ddot{\phi}}{H \dot{\phi}},~~~ \delta_1\equiv4\dot{\xi}H ,~~~\delta_2\equiv\frac{\dot{\delta_1}}{H \delta_1}.
	\ea
In order to attain a precise estimation of the scalar power spectrum,
numerical solving of  the following  Mukhanov-Sasaki (MS) equation is imperative
	\be
	u_k''+\left( k^2-\frac{Z_s''}{Z_s}\right)u_k=0,
	\label{M.S}
	\ee
where the prime stands for derivation with respect to the conformal time $\tau_s=\int\left({c_s}/{a}\right){\rm d}t$
and
	\be
	u_k=Z_s \zeta_k,
	\qquad
	Z_s=a \sqrt{2Q_s c_s},
	\label{ZS}
	\ee
	with
	\begin{align}
	Q_s=\frac{(8\dot{\phi}^2-48 \Mpl^2 H^2+384 H^3 \xi_{,\phi}\dot{\phi})}{(\Mpl^2 H-6 H^2 \xi_{,\phi}\dot{\phi})^2}Q_t^2+12Q_t,\nonumber\\	c_s^2=\frac{1}{Q_s}\left[ \frac{16}{a}\frac{\rd}{\rd t}\left(\frac{a~Q_t^2}{(\Mpl^2 H-6 H^2 \xi_{,\phi}\dot{\phi})}\right)-4c_t^2 Q_t\right],
	\label{cs}
	\end{align}
	%
	%with
	%
	\begin{align}
	Q_t=\frac{1}{4}\left(\Mpl^2-4H\xi_{,\phi}\dot{\phi}\right),\qquad\nonumber\\
	c_t^2=\frac{1}{4Q_t}\left(\Mpl^2-4\xi_{,\phi\phi}\dot{\phi}^2-4\xi_{,\phi}\ddot{\phi}\right).
	\label{ct}
\end{align}
The following criteria are necessary to preclude the  Laplacian and ghost instabilities linked with the scalar and tensor perturbations in the model
	\be
	Q_s>0,\qquad c_s^2>0,\qquad
	Q_t>0,\qquad c_t^2>0.
	\label{avoidinstability}
	\ee
Subsequently, by utilizing the solution of the MS equation (\ref{M.S}), the curvature power spectrum can be computed as
	\be
	{\cal P}_{s}(k)\equiv \frac{k^3}{2\pi^2} \left| \zeta_k (\tau_{s},{ k})\right|^2.\label{powerspectraSUSY}
	\ee
In the regime of the slow-roll conditions ($|\e_{i}|\ll 1$ and $|\delta_{i}|\ll 1$),
the background Eqs.~(\ref{bge1})-(\ref{bge3}) are reduced to
    %simplify
    %	
	\ba
	\label{sre1}
	&& H^2 \simeq \frac{V(\phi)}{3\Mpl^{2}},\\
	\label{sre2}
	&& -2{\Mpl^{2}}\dot{H} \simeq \dot{\phi}^2 + 4\dot{\xi}H^3,\\
	\label{sre3}
	&& 3H\dot{\phi}+V_{,\phi} \simeq- 12\xi_{,\phi}H^4.
	\ea
Also, under slow-roll conditions, the amplitude of the scalar power spectrum ${\cal P}_s(k)$ at the moment of sound horizon crossing is expressed as follows \cite{Kawaguchia:2023,Felice:2011}	
	\be
	{\cal P}_{s}(k)\simeq \frac{H^2}{8\pi^2 Q_s c_s^3}\biggr|_{c_s k=a H}.\label{powerspectraSRSUSY}
	\ee
The observational constraint on the amplitude of curvature power spectrum at the pivot scale $k_{*}=0.05~\rm Mpc^{-1}$ is defined by the Planck team as  ${\cal P}_s(k_{*}) = 2.1\times{10}^{-9}$ \cite{akrami:2020}.
The scalar spectral index $n_s$ in the GB model in terms of the slow-roll parameters, Eq. (\ref{SRH}), is derived as \cite{Jiang:2013}
\ba
n_s-1 \equiv \frac{\rd \ln {\cal P}_s}{\rd \ln k}\biggr|_{c_s k=aH} \simeq -2\e_1-\frac{2\e_1\e_2-\delta_1\delta_2}{2\e_1-\delta_1}.
\label{nsdelta}
\ea
After performing algebraic calculations and using the slow-roll approximation of the background Eqs. (\ref{sre1})-(\ref{sre3}), the slow-roll parameters (\ref{SRH}), can be expressed in terms of the potential and non-minimal coupling GB function as \cite{Jiang:2013,Koh:2014}
\ba
\e_1 &\simeq& \e_V + \frac{2}{3 \Mpl^{2}} V_{,\phi} \xi_{,\phi},\quad\qquad\qquad\qquad\qquad\nonumber\\
\e_2 &\simeq& 4 \e_V - 2 \eta_V -\frac{4}{3 \Mpl^{2}} \frac{V}{V_{,\phi}}\left(\xi_{,\phi} V_{,\phi\phi}+V_{,\phi}\xi_{,\phi\phi} \right),\qquad\quad\nonumber\\
\delta_1 &\simeq& -\frac{4}{9 \Mpl^{4}} \xi_{,\phi}\left(3 \Mpl^{2} V_{,\phi}+\frac{4}{ \Mpl^{2}} V^2 \xi_{,\phi}  \right),\quad\nonumber\\
\delta_2 &\simeq&-\eta_V-\frac{8}{3 \Mpl^{2}}V \xi_{,\phi\phi}-V_{,\phi}\left(\frac{8}{3 \Mpl^{2}}\xi_{,\phi}+\frac{\Mpl^{2} \xi_{,\phi\phi}}{V \xi_{,\phi}} \right).
\label{SRV}
\ea
In which $\e_V \equiv \left( {\Mpl^{2}}/{2}\right) \left( {V_{,\phi}}/{V}\right)^2$ and $\eta_V \equiv \Mpl^{2} \left(V_{,\phi\phi}/V\right)$ are the potential slow-roll parameters in the standard inflation. Replacing the relations (\ref{SRV}) into Eq. (\ref{nsdelta}), one can get the scalar spectral index $n_s$ as
\ba
n_s-1 \simeq -6\e_{V} + 2 \eta_{V} + \frac{4}{3 \Mpl^{2}} V \xi_{,\phi} \left(\frac{V_{,\phi}}{V}+\frac{2 \xi_{,\phi\phi}}{\xi_{,\phi}}\right).
\label{ns}
\ea
Note that,
the equivalent counterpart of Eq.~(\ref{ns}) in the standard inflationary model is recovered for $\xi=0$.
The observational constraint on the scalar spectral index, imposed by  Planck 2018 TT, TE, EE + lowE + lensing + BK18 + BAO \cite{akrami:2020}, is $n_s=0.9649\pm 0.0042$ $(68\,\%\,\rm CL)$.

Identically, the tensor power spectrum is calculated using the subsequent MS equation \cite{Kawaguchia:2023,Felice:2011,Kawai:2021}
	\be
	u_k''+\left( k^2-\frac{Z_t''}{Z_t}\right)u_k=0,
	\label{M.St}
	\ee
where prime denotes variation with respect to the conformal time $\tau_t=\int\left({c_t}/{a}\right){\rm d}t$.
Also

	\be
	u_k=Z_t \zeta_t,\qquad Z_t=a \sqrt{2Q_t c_t}.
	\label{ZT}
	\ee
	Solving the MS equation (\ref{M.St}) yields the tensor power spectrum for each mode $u_k$ in the following manner
	\be
	{\cal P}_{t}(k)\equiv 4\frac{k^3}{2\pi^2} \left| \zeta_t (\tau_{t},{ k})\right|^2.\label{powerspectraZetaT}
	\ee
Additionally, under the slow-roll condition at the horizon crossing time, the tensor power spectrum ${\cal P}_t(k)$ is determined as follows \cite{Kawaguchia:2023,Ps&PtHorndeski,Odintsov:2020nsr,Felice:2011,Jiang:2013,Koh:2014}	
	\be
	{\cal P}_{t}(k)\simeq \frac{H^2}{2\pi^2 Q_t c_t^3}\biggr|_{c_t k=a H}.\label{powerspectraZetaTSR}
	\ee
The tensor to scalar ratio $r$ in the GB model in terms of the slow-roll parameters (\ref{SRH}) is derived as
	\ba
	r \equiv \frac{{\cal P}_t}{{\cal P}_s}\simeq 8\left(2\e_1-\delta_1\right)\label{r0}.
	\ea
By using the slow-roll parameters (\ref{SRV}), the tensor to scalar ratio $r$ reads
	\ba
	r \simeq 16\e_{V} + \frac{64}{3 \Mpl^{2}} \xi_{,\phi} V\left(\frac{V_{,\phi}}{V}+\frac{2}{3 \Mpl^{4}} \xi_{,\phi} V\right)\label{r}.
	\ea
The estimated upper limit on the tensor to scalar ratio, reported by Planck 2018 TT, TE, EE + lowE + lensing + BK18 + BAO measurements, is $r<0.036~$ $(95\%\rm CL)$ \cite{akrami:2020,Ade:2021,Campeti:2022}.
Notably, the standard inflation results can be achieved by choosing $\xi=0$ in Eq.~(\ref{r}).

	%%========================section 3 =========================
	\section{Amplification of the curvature power spectrum}
	\label{sec3}
PBHs can be generated through various methods, one of them is utilizing the USR phase. During this phase, the amplitude of the scalar power spectrum at small scales increases, leading to the formation of PBHs. In the GB modified gravity model,
by choosing a suitable coupling function and fine-tuning of the free parameters of model, the curvature power spectrum can be enhanced significantly.
In the following, we introduce a proper coupling function to achieve this goal in our model.
The adjustment of model parameters serves two primary purposes: firstly, ensuring the model consistency with Planck observational measurements at the CMB scale and secondly, seven-order of magnitude increase in the scalar power spectrum at small scales. To achieve both objectives, we employ a two-parted coupling function as follows

    \be
    \label{eqxi}
    \xi(\phi)=\xi_{\rm I}(\phi) \left[1+\xi_{\rm II}(\phi)\right],
    \ee	
where
 \be
    \label{eqxi12}
  \xi_{\rm I}(\phi)= M \exp\left(-\beta~\phi \right),  \ \ \ \ \xi_{\rm II}(\phi)= \xi_0\tanh\left[\xi_1(\phi-\phi_c)\right].
    \ee
The first part of the coupling function, $\xi_{\rm I}(\phi)$, is a dilatonic continuous function to make the model consistent with the observational data on the CMB scale. The parameters $M$ and $\beta [\Mpl^{-1}]$ of $\xi_{\rm I}(\phi)$ are used to adjust the model with the CMB anisotropies observations at the pivot scale. The $\xi_{\rm II}(\phi)$ function of the second part of the coupling function Eq.~(\ref{eqxi}), is chosen phenomenologically to make the inflaton field decelerate on small scales, far from the CMB scale, to create a peak in the curvature power spectrum \cite{Kawai:2021,Kawaguchia:2023,Zhang:2022}. Besides, on the peak position, $\xi_{\rm I}(\phi)$ plays an amplifying role. The $\xi_{\rm II}(\phi)$ is a step-like function with sharp change around $\phi=\phi_c$ and constant asymptotic values far from $\phi_c$ as $\xi_{\rm II}(\phi)=-\xi_0$ for $\phi\ll\phi_c$ and $\xi_{\rm II}(\phi)=\xi_0$ for $\phi\gg\phi_c$. Hence, we can achieve a particular feature on the scales around $\phi_c$ to decelerate the inflaton and produce a peak in the curvature power spectrum without affecting on the farther scales around the CMB and end of inflation. In the following, the position, width and height of the generated peak in the scalar power spectrum are, respectively, controlled by $\phi_{c}[\Mpl]$, $\xi_{1}[\Mpl^{-1}]$ and $\xi_{0}$.
Note that $M$ and $\xi_{0}$ are dimensionless.

\begin{figure}[H]
	\begin{minipage}[b]{1\textwidth}
		\subfigure[\label{PHISUSY} ]{\includegraphics[width=0.45\textwidth]%
			{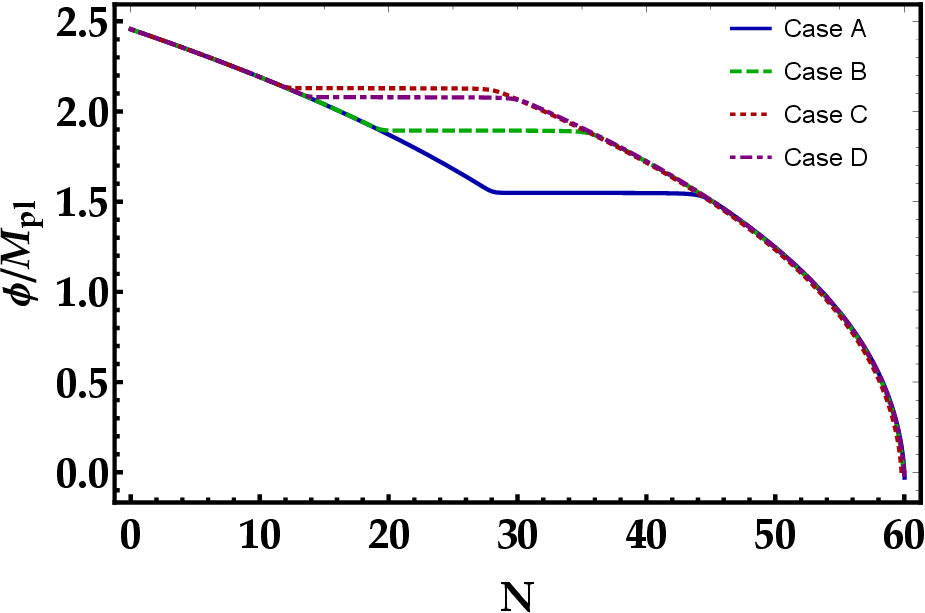}}
		\hspace{.1cm}
		\subfigure[\label{HSUSY}]{\includegraphics[width=.45\textwidth]%
			{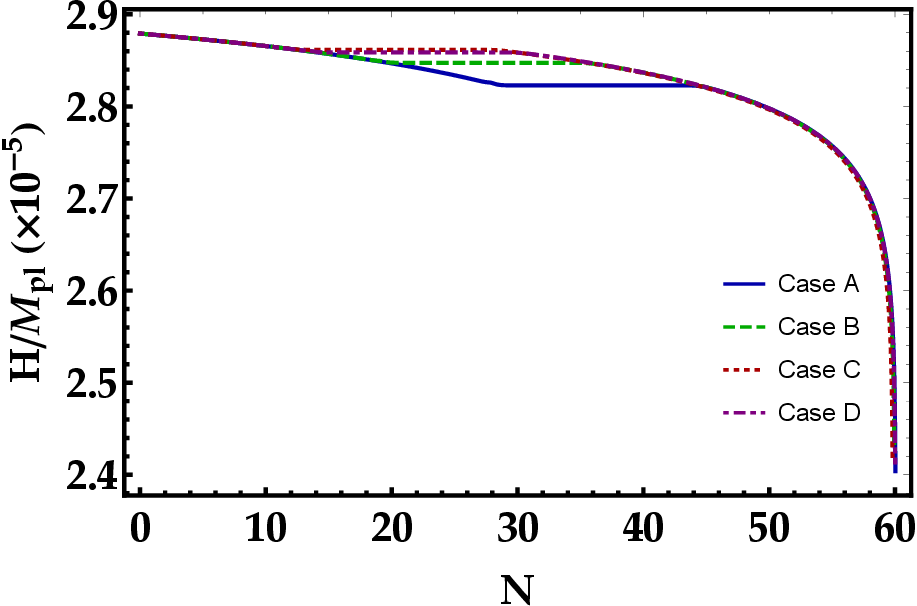}}
		\subfigure[\label{EPSILONSUSY}]{\includegraphics[width=.47\textwidth]%
			{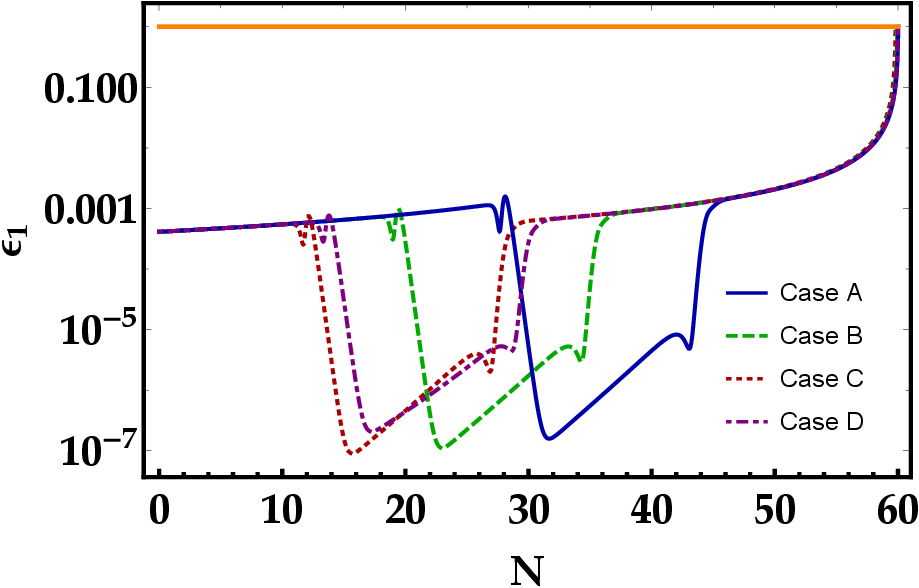}}
		\hspace{.6cm}
		\subfigure[\label{ETASUSY}]{\includegraphics[width=.45\textwidth]%
			{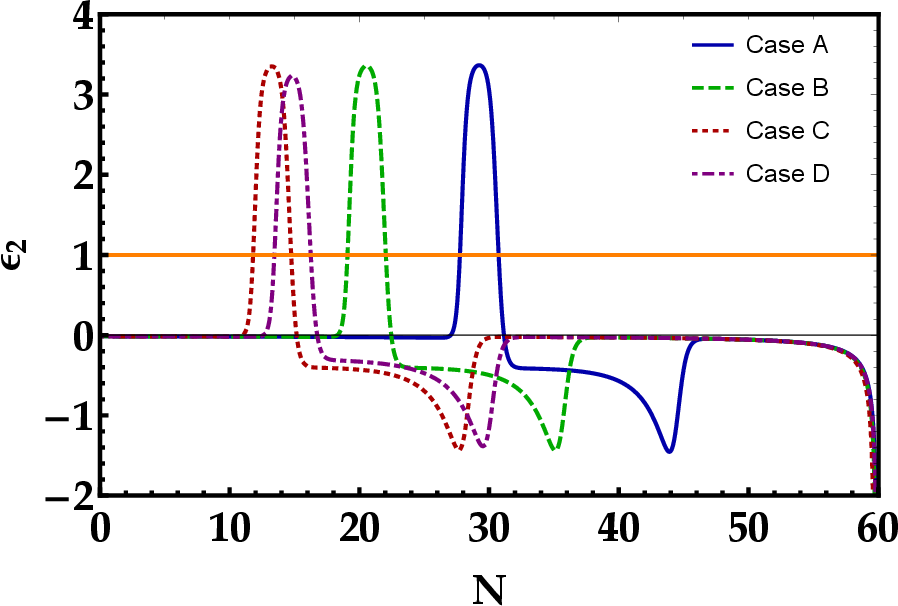}}
		\subfigure[\label{CSSUSY}]{\includegraphics[width=.47\textwidth]%
			{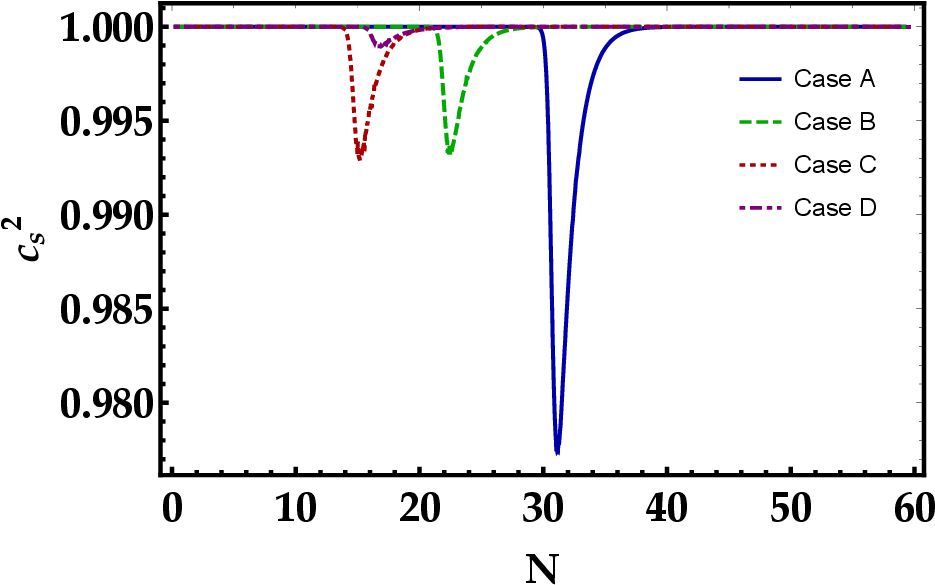}}
		\hspace{.6cm}
		\subfigure[\label{CtSUSY}]{\includegraphics[width=.48\textwidth]%
			{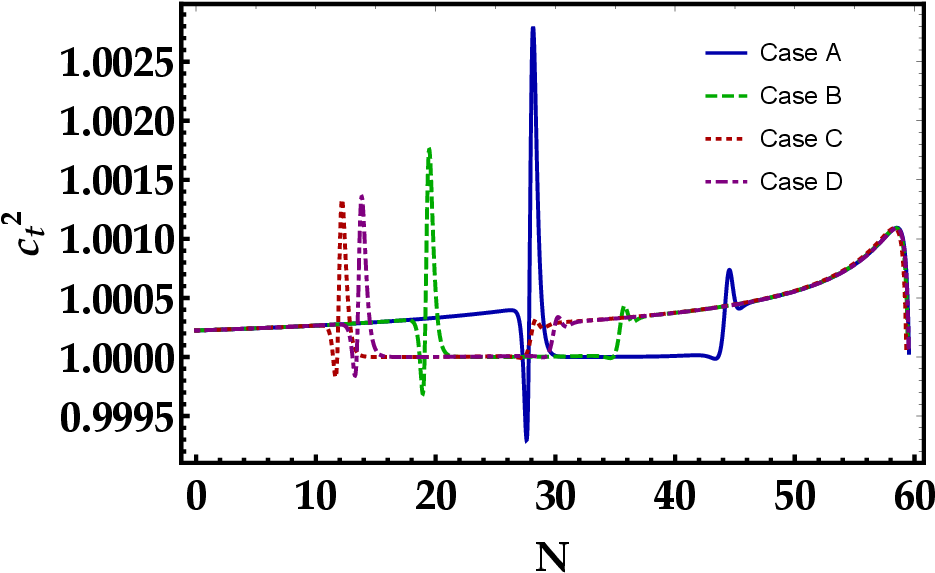}}
		
	\end{minipage}
		\caption{Evolution of (a) the scalar filed $\phi$, (b) the Hubble parameter, (c) the first slow-roll parameter $\e_1$, (d) the second slow-roll parameter $\e_2$, (e) $c_s^{2}$, (f) $c_t^{2}$ versus the $e$-folds number $N$ for cases of Table~\ref{tab1}.
		%The blue, green, red and purple lines are corresponding to $\rm{Case~}A$, $\rm{Case~}B$, $\rm{Case~}C$ and $\rm{Case~}D$, respectively.
		The cases A, B, C, and D are respectively shown by blue, green, red and purple lines.}
		\label{Fig1}
	\end{figure}

To justify the behavior of universe, it is preferred to use potentials rooted in particle physics. The spontaneously broken supersymmetric (SB SUSY) potential is one of the inflationary potentials that originates from particle physics~\cite{Dvali:1994,Copeland:1994,Linde:1994}.
The SB SUSY potential is ruled out of the observational scope of Planck 2018 in the standard inflationary model~\cite{akrami:2020}. This characteristic prompted us to reassess it within the context of
the GB gravity theory to enhance the compatibility of the  model with observations.
The inflationary low-scale SB SUSY potential can be considered as follows ~\cite{Dvali:1994,Copeland:1994,Linde:1994}
     \be
    \label{eqv}
    V(\phi)=V_0 \left[1+\alpha \ln \left(\frac{\phi}{\Mpl}\right)\right],
    \ee
where, $V_{0}[\Mpl^{4}]$
could be fixed by the Planck measurements of the scalar power spectrum  ${\cal P}_s(k_{*}) = 2.1\times{10}^{-9}$ at the pivot scale $k_{*}=0.05~\rm Mpc^{-1}$.
Also, the parameter $\alpha$ is dimensionless and takes values in the range  $-2.5\leq\log\alpha\leq 1$.
It is worth noting that the USR phase can be generated in the vicinity of a non-trivial fixed point and this point can be obtained in the presence of the GB term. Using the conditions $\dot{H}=\dot{\phi}=\ddot{\phi}=0$ at the fixed point and applying Eq.~(\ref{bge3}), the following criterion at $\phi = \phi_c$ can be obtained \cite{Kawai:2021,Kawaguchia:2023,Zhang:2022}
	\be
	\Bigg[ V_{,\phi}+\frac{4 \xi_{,\phi}{V(\phi)}^2}{3\Mpl^4}\Bigg]
	\Biggr|_{\phi=\phi_c}
	=0,
	\label{EqC}
	\ee
which results in the following equation for the low-scale SB SUSY potential (\ref{eqv})
	\be
	\xi_{0}=\frac{\beta}{\xi_1}-\frac{3\Mpl^4\alpha \exp \left( \beta \phi_{c}\right)}{4 M V_{0} \xi_1 \phi_{c}\left[1+\alpha \ln\left(\frac{\phi_{c}}{\Mpl}\right)\right]^2}.
	\label{EqCSUSY}
	\ee
By employing Eq.~(\ref{EqCSUSY}), the parameters are defined and fine-tuned to achieve the objectives of this study. Note that, regarding the condition $\dot{\phi}\bigr|_{\phi=\phi_c}=0$, we have
$\dot{V}\bigr|_{\phi=\phi_c}=V_{,\phi}\dot{\phi}\bigr|_{\phi=\phi_c}=0$, and the $\phi_c$ can be considered as the fixed point of the potential (\ref{eqv}).
In order to define the evolution of $H$ and $\phi$ in terms of the $e$-folds number $N$, using $dN = Hdt$, we solve Eqs.~(\ref{bge2}) and (\ref{bge3}) numerically.
Here, the slow-roll approximations, outlined in Eqs. (\ref{sre1}) and (\ref{sre3}), have been utilized to determine the required initial values for solving the background equations.
In addition, the inflation epoch in our model lasts for 60 $e$-folds number, which is started from $N_{*}=0$ (the horizon crossing $e$-fold by the CMB scale).
For all cases in this study, the $e$-folds number at the end of inflation $N_{\rm end}$ is determined by $\e_1 = 1$, as illustrated in Fig.~\ref{EPSILONSUSY}.
In the vicinity of the fixed point, when the inflaton enters the USR phase $\e_2 > 1$ and slow-roll approximation becomes invalid, as illustrated in Fig.~\ref{ETASUSY}.
Subsequently, the value of the scalar power spectrum cannot be determined by Eq.~(\ref{powerspectraSRSUSY}), which is estimated under the slow-roll conditions.

Consequently, we must  solve the MS equation (\ref{M.S}), numerically, to calculate the precise values of the scalar power spectrum and PBH abundance in the presence of a low-scale SB SUSY potential.
Utilizing Eq.~(\ref{EqCSUSY}), we can
compute
the value of $\xi_0$
in terms of
$\beta$, $\xi_1$, $\alpha$, $M$, $V_0$ and $\phi_c$. However, a slight fine-tuning of the estimated $\xi_0$ value is necessary to attain  sufficient abundance of PBHs.	
In Fig.~\ref{Fig2}, using the solution of the MS equation (\ref{M.S}), the evolution of ${\cal P}_{s}$ is depicted as a function of the co-moving wavenumber $k$.
The adjusted values of four parameter sets are listed in Table~\ref{tab1}. The final column of this table displays the value of $\xi_0$ estimated using Eq.~(\ref{EqC}).
The quantities  of the scalar spectral index $n_s$ and the tensor to scalar ratio $r$ at the pivot scale have been estimated by utilizing, respectively, Eqs.~(\ref{ns}) and (\ref{r}) for all cases of the model.
Our calculations yield $n_s=0.970$, that align with the $68\,\%\,\rm CL$ of the Planck 2018 data TT, TE, EE + lowE + lensing + BK18 + BAO \cite{akrami:2020,Campeti:2022} and $r=0.005$ align with the BICEP/Keck 2018 data $95\,\%\,\rm CL$ \cite{akrami:2020,Ade:2021}.

\begin{figure}[H]
		\centering
\hspace*{-1cm}
\includegraphics[scale=0.55]{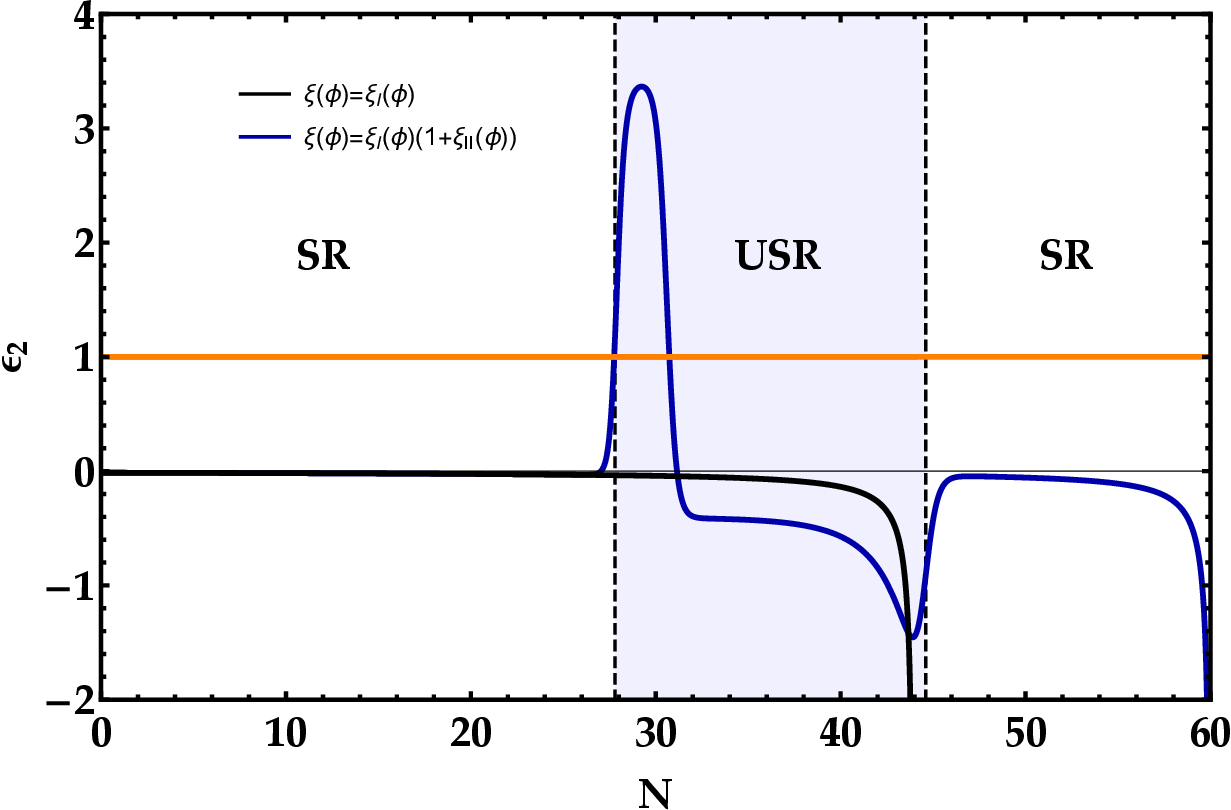}
\caption{Evolution of the second slow-roll parameter $\epsilon_2$ with respect to the $e$-fold number $N$ for case A of the model with $\xi(\phi)=\xi_{\rm I}(\phi)$ (black curve) and $\xi(\phi)=\xi_{\rm I}(\phi)[1+\xi_{\rm II}(\phi)]$ (blue curve).}
		\label{sr-usr}
	\end{figure}

Furthermore, Figs. \ref{PHISUSY} and \ref{HSUSY} illustrate the evolution of the scalar field $\phi$ and the Hubble parameter $H$, respectively.
The plateau eras observed in both figures correspond to the USR regime, characterized by a severe decrease in the inflaton velocity (i.e. $\dot{\phi}\rightarrow 0$) and significant amplification of primordial curvature perturbation (see Fig. \ref{Fig2}).
It should be noted that, the first slow-roll parameter $\epsilon_1$ stays below unity during the USR phase (see Fig. \ref{EPSILONSUSY}),
while the second slow-roll parameter $\epsilon_2$ increases and surpasses one, signifying
the failure of the slow-roll approximation. Note that, in our model, two transitions occur as SR$\rightarrow$USR$\rightarrow$SR. The first transition  (SR$\rightarrow$USR) occurs at $\epsilon_2>1$ and the second one (USR$\rightarrow$SR) takes place  at  $\epsilon_2<-1$ (see Fig. \ref{ETASUSY} and also Fig. \ref{sr-usr}).
Additionally,
in Figs. \ref{CSSUSY} and \ref{CtSUSY}, variation of $c_s^{2}$ and $c_t^{2}$ versus the $e$-folds number are depicted.
These plots confirm
that the present model
adheres to the stability conditions outlined in Eq.~(\ref{avoidinstability}).
Lastly, the computed values of the scalar power spectra at the peak positions
${\cal P}_{s}^\text{peak}$ are presented in Table \ref{tab2}.

 \begin{table}[H]
	\centering
	\caption{The parameter sets of the model. For all sets we take $M=3.91\times 10^{7}$, $\beta=0.09~\Mpl^{-1}$, $V_0=2.30\times 10^{-9}$ $\Mpl^{4}$, $\alpha=0.09$, $\phi_{*} = 2.45~\Mpl$, $n_s=0.970$ and $r=0.005$. The quantities of $\phi_{*} $, $n_s$ and $r$
		are calculated at the horizon passing moment $N_{*} = 0$ of the pivot scale.}
	\begin{tabular}{ccccc}
	\hline
	Sets \quad &\quad $\phi_{c}$$\left[ \Mpl\right]$ \quad & \quad $\xi_{1}$$\left[\Mpl^{-1}\right]$\quad &$\xi_{0}$\quad & $\xi_{0} \left(\rm From~Eq.~(\ref{EqCSUSY})\right)$\quad  \\ [0.5ex]
	\hline
	\hline
	$\rm{Case~}A$ \quad &\quad $1.55$ \quad &\quad $60.5$ \quad &\quad $-7.0717\times10^{-3}$&\quad $-7.0229\times10^{-3}$ \quad\\[0.5ex]
	\hline
	$\rm{Case~}B$ \quad &\quad $1.89$ \quad &\quad $78.2$\quad &\quad $-4.2445\times10^{-3}$&\quad $-4.2161\times10^{-3}$ \quad\\[0.5ex]
	\hline
	$\rm{Case~}C$ \quad &\quad $2.13$ \quad &\quad $91.5$\quad &\quad $-3.1236\times10^{-3}$&\quad $-3.1027\times10^{-3}$ \quad\\[0.5ex]
	\hline
	$\rm{Case~}D$ \quad &\quad $2.08$ \quad &\quad $85.0$\quad &\quad $-3.4565\times10^{-3}$&\quad $-3.4434\times10^{-3}$ \quad\\[0.5ex]
	\hline
	\end{tabular}
	\label{tab1}
	\end{table}

	\begin{table}[H]
		\centering
		\caption{The values of $k_{\text{peak}}$, ${\cal P}_{s}^\text{peak}$, $M_{\text{PBH}}^{\text{peak}}$ and $f_{\text{PBH}}^{\text{peak}}$ for the cases listed in Table \ref{tab1}.}
		\begin{tabular}{ccccccc}
			\hline
			Sets \quad & \quad$k_{\text{peak}}/\text{\rm Mpc}^{-1}$ \quad &\quad ${\cal P}_{s}^\text{peak}$ \quad & \quad  $M_{\text{PBH}}^{\text{peak}}/M_{\odot}$ \quad & \quad $f_{\text{PBH}}^{\text{peak}}$\\ [0.5ex]
			\hline
			\hline
			$\rm{Case~}A$ \qquad &\quad $3.8982\times10^{12}$ \quad &\quad $0.0312$ \quad &\quad $1.95\times 10^{-13}$ \quad &$0.9980$ \\[0.5ex]
			\hline
			$\rm{Case~}B$ \quad &\quad $7.2027\times10^{8}$ \quad &\quad $0.0412$ \quad &\quad $6.65\times10^{-6}$ \quad & $0.0546$ \\[0.5ex]
			\hline
			$\rm{Case~}C$ \quad &\quad $4.8902\times10^{5}$ \quad &\quad $0.0494$ \quad &\quad $14.38$ \quad &$0.0023$\\[0.5ex]
			\hline
			$\rm{Case~}D$ \quad &\quad $2.1858\times10^{6}$ \quad &\quad $0.0107$ \quad &\quad $0.583$ \quad &$\sim 0$\\[0.5ex]
			\hline
		\end{tabular}
		\label{tab2}
	\end{table}

 In our GB model, the USR phase is controlled by the non-minimal coupling function (\ref{eqxi}). In fact, the second term of the coupling function $\xi(\phi)$ in Eq.~(\ref{eqxi}), through the $\xi_{\rm II}(\phi)$ function, is responsible to produce the USR phase in the GB model. In Fig. \ref{sr-usr}, the second slow-roll parameter $\epsilon_2$ of the model is shown for case A, for instance, when the coupling function $\xi(\phi)$ contains only $\xi_{\rm I}(\phi)$ function (black curve) and when it takes its complete form Eq.~(\ref{eqxi}) with $\xi_{\rm II}(\phi)$ (blue curve). This figure shows that, if $\xi(\phi)$ is composed of the only $\xi_{\rm I}(\phi)$ function, there is no violation of the slow-roll condition to produce the USR phase in the inflationary era to generate PBHs. In this case, the inflationary era lasts for almost $44$ $e$-folds. Also, it can be seen from Fig. \ref{sr-usr} that, when $\xi(\phi)$ contains the $\xi_{\rm II}(\phi)$ function, the $\epsilon_2$ plot violates the slow-roll condition and the USR phase is added to the inflationary era for around $16$ $e$-folds number to make the inflation last for $60$ $e$-folds number. So, this figure shows how the particular form of the non-minimal coupling Eqs.~(\ref{eqxi}) and (\ref{eqxi12}) in our GB model can drive a SR—USR—SR phase to get a peak in the curvature perturbation to produce PBHs.

\begin{figure}[H]
		\centering
\hspace*{-1cm}
\includegraphics[scale=0.55]{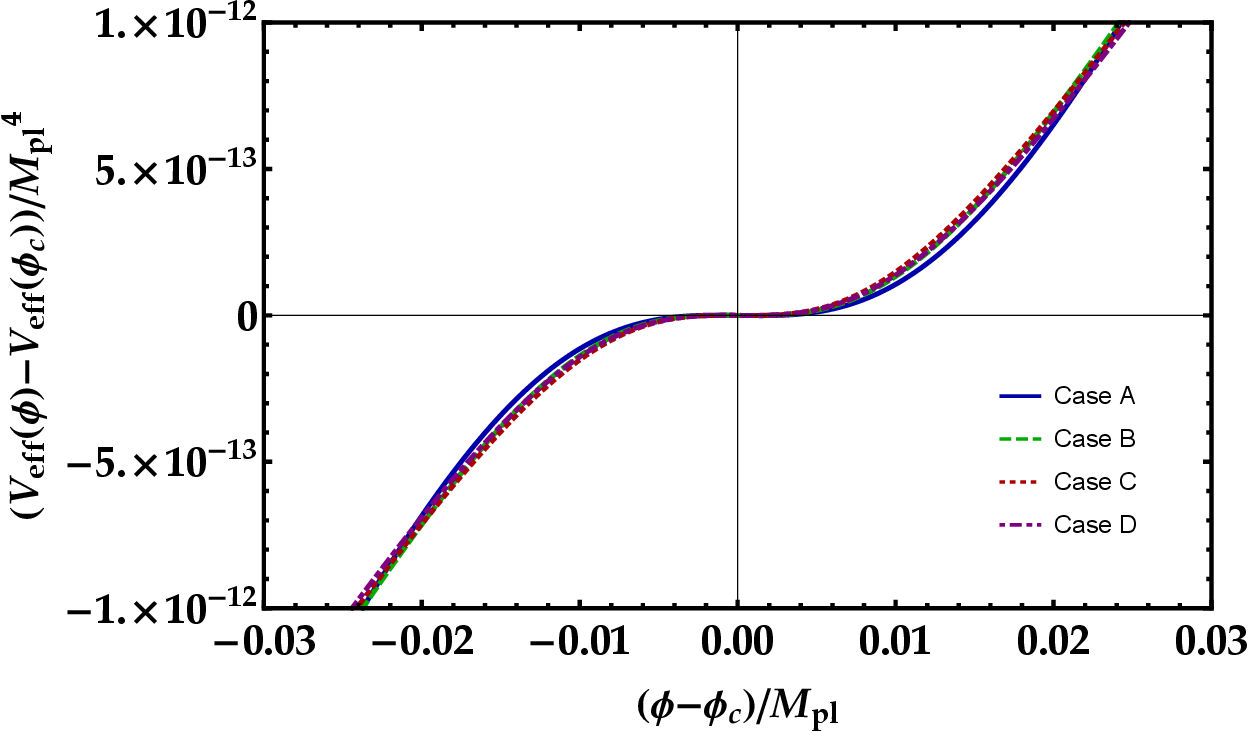}
\caption{The effective potential, Eq.~(\ref{veff}), with respect to the scalar field for all cases of the model in the Einstein frame.}
		\label{vefff}
	\end{figure}

Note that, in consistency with the Einstein framework, the equation of motion Eq.~(\ref{bge3}) can be rewritten as follows
	\ba
	\label{bge4}
    \ddot{\phi}+3H\dot{\phi} + V_{\rm{eff},\phi}=0 ,
	\ea
where
	\ba
	\label{veff,phi}
   V_{\rm{eff},\phi}=V_{,\phi}+12\xi_{,\phi}H^2\big(\dot{H}+H^2\big).
	\ea
Using the evolutions of $\phi$ and $H$ versus the $e-$fold number $N$ in Eq.~(\ref{veff,phi}), we can go to the Einstein frame and calculate the effective potential numerically, as follows
\ba
	\label{veff}
   V_{\rm{eff}}(N)=\int{\Big[V_{,\phi}+12\xi_{,\phi}H^3\big(H_{,N}+H\big)\Big]\phi_{,N}dN,}
	\ea
where $(,N)$ stands for derivative with respect to $N$. In this regard, the USR phase to generate PBHs can be introduced around the inflection point of the effective potential Eq.~(\ref{veff}), as it can be seen from Fig. \ref{vefff}. This figure shows that the inflection points of the effective potential for all cases of the model occur at $\phi=\phi_c$ tabulated in Table \ref{tab1}.
	
\subsection{Swampland's criteria}
\label{swampland}
\begin{figure}[H]
		\begin{minipage}[b]{1\textwidth}
			\centering
			\subfigure[\label{SwapPhi}]{\includegraphics[width=.48\textwidth]%
				{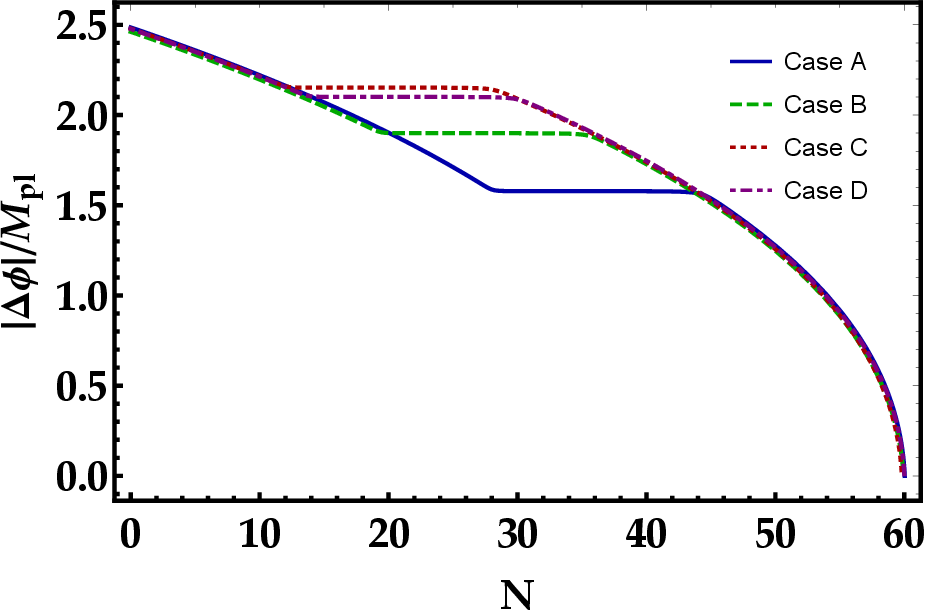}}\\
			\hspace{.1cm}	
			\centering
			\subfigure[\label{SwapV}]{\includegraphics[width=.48\textwidth]%
				{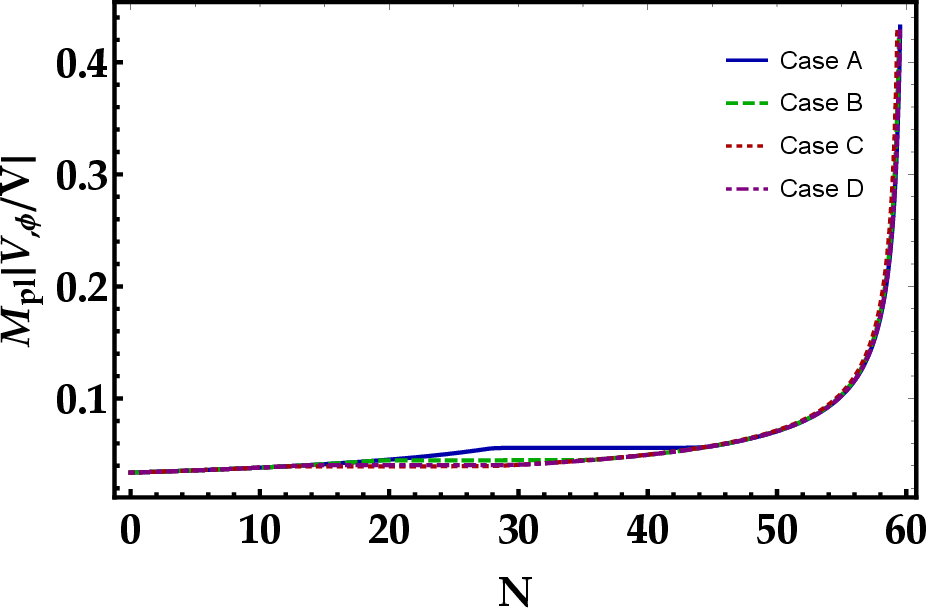}}
\centering
			\subfigure[\label{SwapVV}]{\includegraphics[width=.48\textwidth]%
				{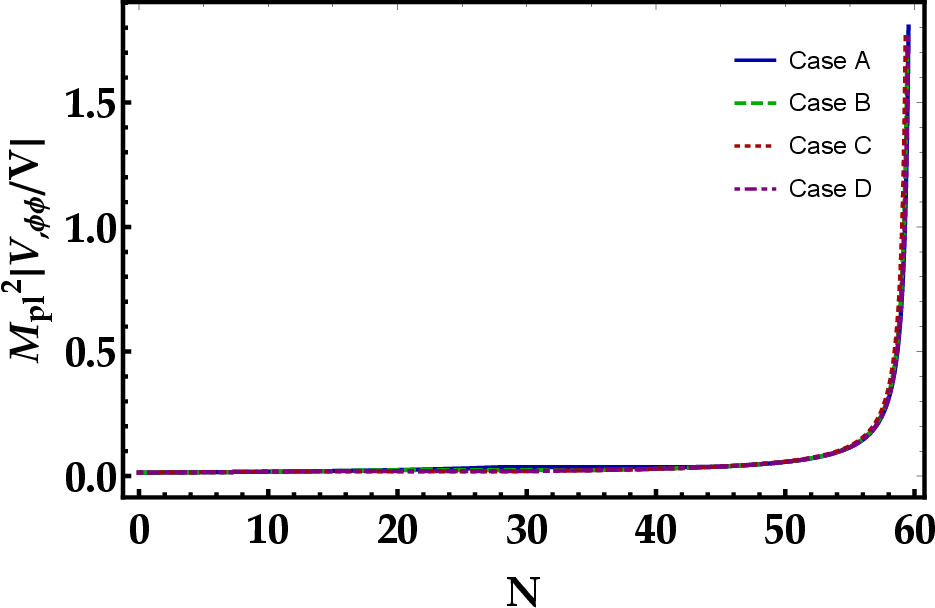}}
				
		\end{minipage}
		\caption{Validity of the swampland criteria for our model for (a) $\Delta\phi= \phi(N) - \phi(N_{\rm end})$,  (b) $\Mpl|V_{,\phi}/V|$, and (c) $\Mpl^2|V_{,\phi\phi}/V|$ versus the $e$-fold number $N$.
			The cases A, B, C, and D are, respectively, depicted by blue, green, red and purple lines.}
		\label{Swaps}
	\end{figure}

Initially, we will introduce the swampland criteria and highlight the importance of incorporating these criteria in the analysis of inflationary models. The swampland criteria emerge from the string theory and encompass two principal theoretical conjectures known as distance and de Sitter conjectures.
The distance conjecture is defined as follows
\begin{equation}
\frac{|\Delta \phi|}{\Mpl} \lesssim c_1,
\label{swap1}
\end{equation}
where $c_1$ is a constant parameters. Recent studies suggest that $c_1 \sim {\cal O}(1)$.
Our estimations indicate that in the present model, $|\Delta \phi|/\Mpl \sim {\cal O}(1)$, confirming the distance conjecture of the swampland criteria (note Fig. \ref{SwapPhi}) \cite{akrami:2020,Agazie1:2023,Agazie2:2023,Agazie3:2023,Agazie4:2023}. 

In addition, we focus on the refined de Sitter
swampland criterion given by \cite{Garg:2019}
\begin{equation}
\Mpl \left|\frac{V_{,\phi}}{V}\right| \gtrsim c_2,\qquad \text{or}\qquad \Mpl^2 \left|\frac{V_{,\phi\phi}}{V}\right| \gtrsim c_2,
\label{swap2}
\end{equation}
where $c_2$ is a constant parameter of order ${\cal O}$(0.1-1).
It's noteworthy that the standard inflationary model, characterized by $\epsilon_{V},\eta_{V} \ll 1$, is in stark contrast to the de Sitter conjecture.
In the standard inflationary model, the ratio of tensor to scalar perturbations is estimated as $r=16\e_V$. Consequently, considering the definition of $\e_V$ and the de-Sitter conjecture result in $r \gtrsim 8 c_2^2$. Hence, taking the parameter $c_2\sim{\cal O}(0.1)$ yields $r\sim 0.08$, which is inconsistent with Planck measurements at the CMB scale.
This discrepancy serves as a compelling motivation for investigating the compatibility of our model with the swampland conjectures.
Figure \ref{Swaps} demonstrates the fulfillment of the de Sitter conjecture, where $\Mpl|V_{,\phi}/V|$ is of the order of ${\cal O}(0.1)$.
Therefore, our model exhibits acceptable agreement with both the distance and refined de Sitter conjectures of the swampland criteria.

\subsection{Results for ${\cal P}_{s}$ and ${\cal P}_{t}$}
\label{Ps&Pt}

\begin{figure}[H]
		\centering
		\hspace*{-1cm}
		\includegraphics[scale=0.55]{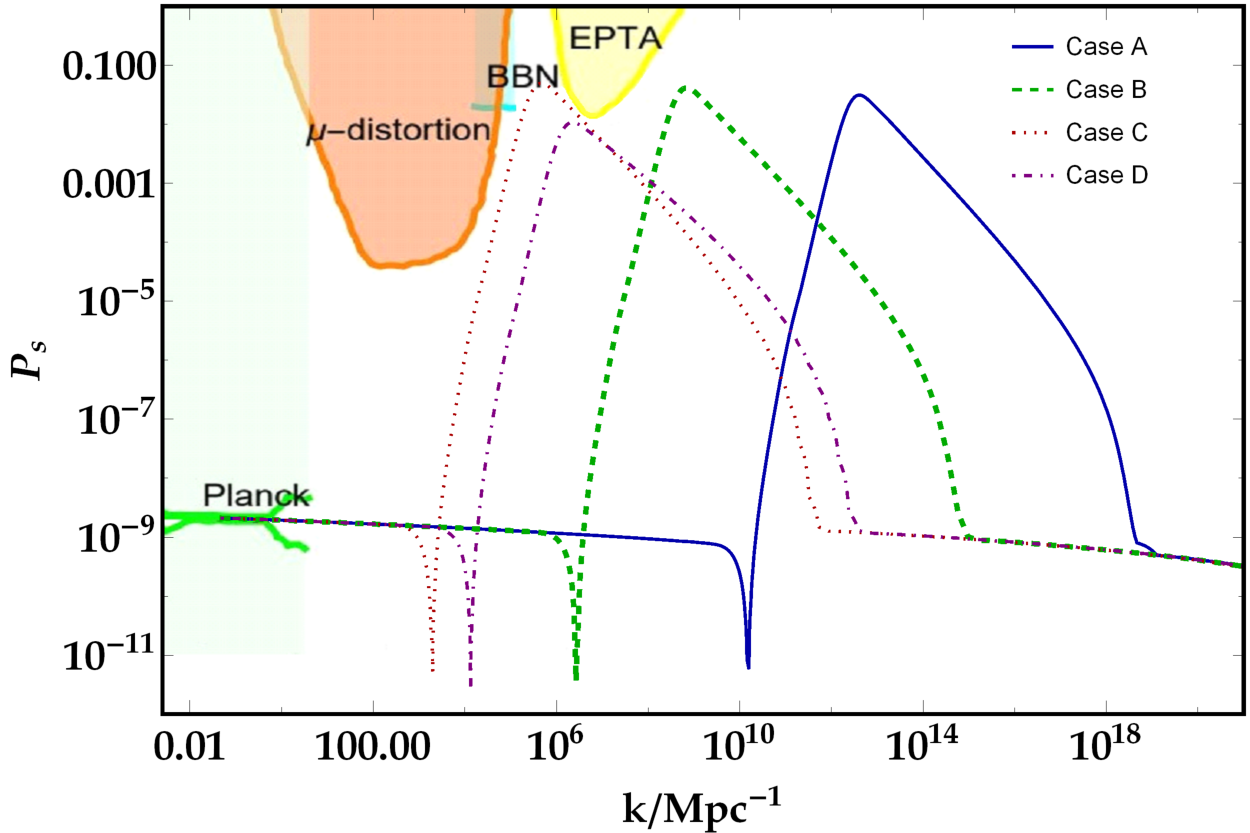}
		\caption{%
			The scalar power spectrum ${\cal P}_{s}$  as a function of the wavenumber $k$ for each parameter set of Table~\ref{tab1}.
		The cases A, B, C, and D are respectively shown by blue, green, red and purple curves.
		The CMB observations exclude the light-green shaded region \cite{akrami:2020}. The orange region shows the $\mu$-distortion of CMB \cite{Fixsen:1996}. The cyan area represents the effect on the ratio between neutron and proton during the big bang nucleosynthesis (BBN) \cite{Inomata:2016}. The EPTA observations constrain the yellow area \cite{Inomata:2019-a}. }
		\label{Fig2}
	\end{figure}

\begin{figure}[H]
		\centering
		\hspace*{-1cm}
		\includegraphics[scale=0.55]{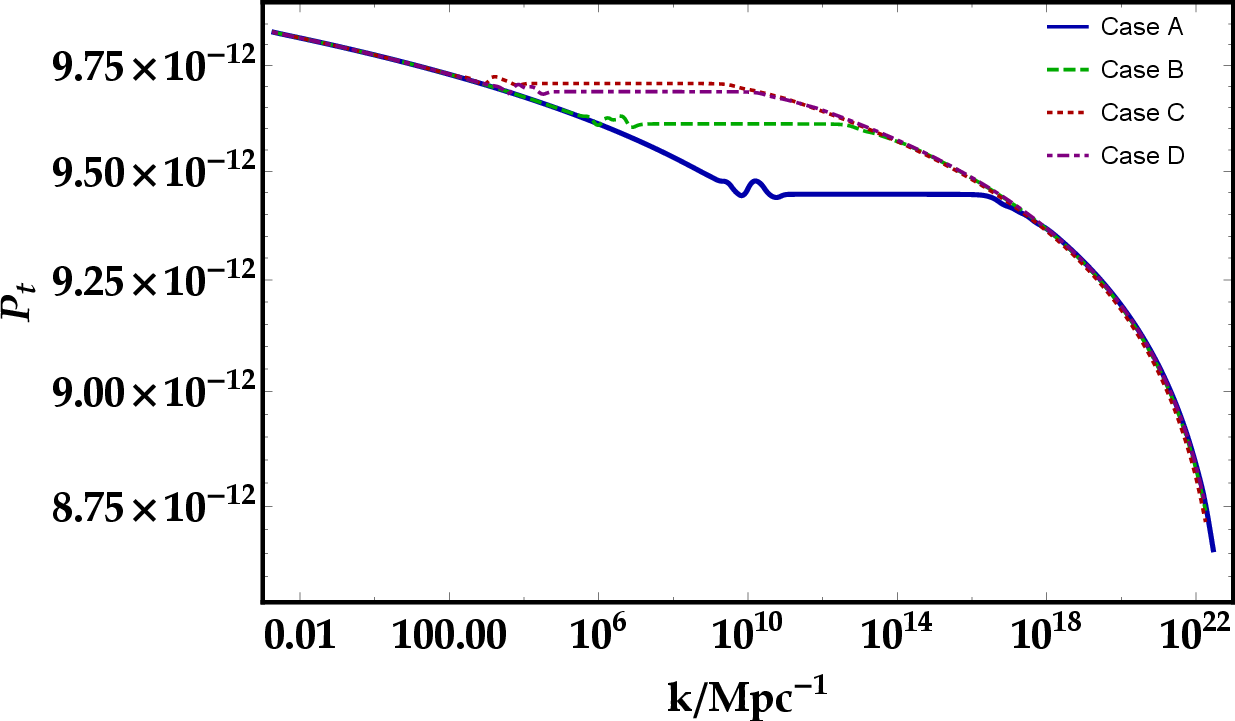}
		\caption{%
			The tensor power spectrum ${\cal P}_{t}$ versus the wavenumber $k$ for each parameter set of Table~\ref{tab1}.
			The cases A, B, C, and D are respectively shown by blue, green, red and purple curves.}
		\label{Fig3}
	\end{figure}

As mentioned previously, after solving the MS Eq.~(\ref{M.S}), the value of the scalar power spectrum is determined. Furthermore, in Fig.~\ref{Fig2}, we depict the evolution of ${\cal P}_{s}$ for all parameter sets listed in Table \ref{tab1}. The curves in Fig.~\ref{Fig2} are generated based on the mathematical calculations mentioned earlier and utilizing Eq.~(\ref{powerspectraSUSY}). The results demonstrate that the scalar power spectrum aligns with the CMB measurements at the pivot scale. Additionally, ${\cal P}_{s}$ exhibits a significant enhancement by approximately seven orders of magnitude at small scales for PBH formation.
Similarly, by solving Eq.~(\ref{M.St}), the precise value of the tensor power spectrum ${\cal P}_{t}$ is obtained for each parameter set of Table \ref{tab1}. The results  are depicted in Fig.~\ref{Fig3}, for all parameter sets. It should be noted that the value of the tensor to scalar ratio $r$, which is obtained from the exact solutions of Eqs. (\ref{powerspectraSUSY}) and (\ref{powerspectraZetaT}), on the CMB scale is consistent with the value obtained from the slow-roll approximation Eq. (\ref{r}).
It should be noted that, because of the constancy of the Hubble parameter in the USR domain, the plateau zone is formed in the evolution of ${\cal P}_{t}$ (see Fig. \ref{HSUSY}).
Furthermore, the minor fluctuations observed in ${\cal P}_{t}$ stem from the oscillatory behavior of $c_t$ (see Fig. \ref{CtSUSY}).	
%

	%%%%%%%%%%%%%%%%%%%%%%%%%%%%%%%%%%%%%%%%%%
	\section{The Abundance of PBHs}
	\label{sec4}
\begin{figure}[h]
	\centering
	\hspace*{-1cm}
	\includegraphics[scale=0.55]{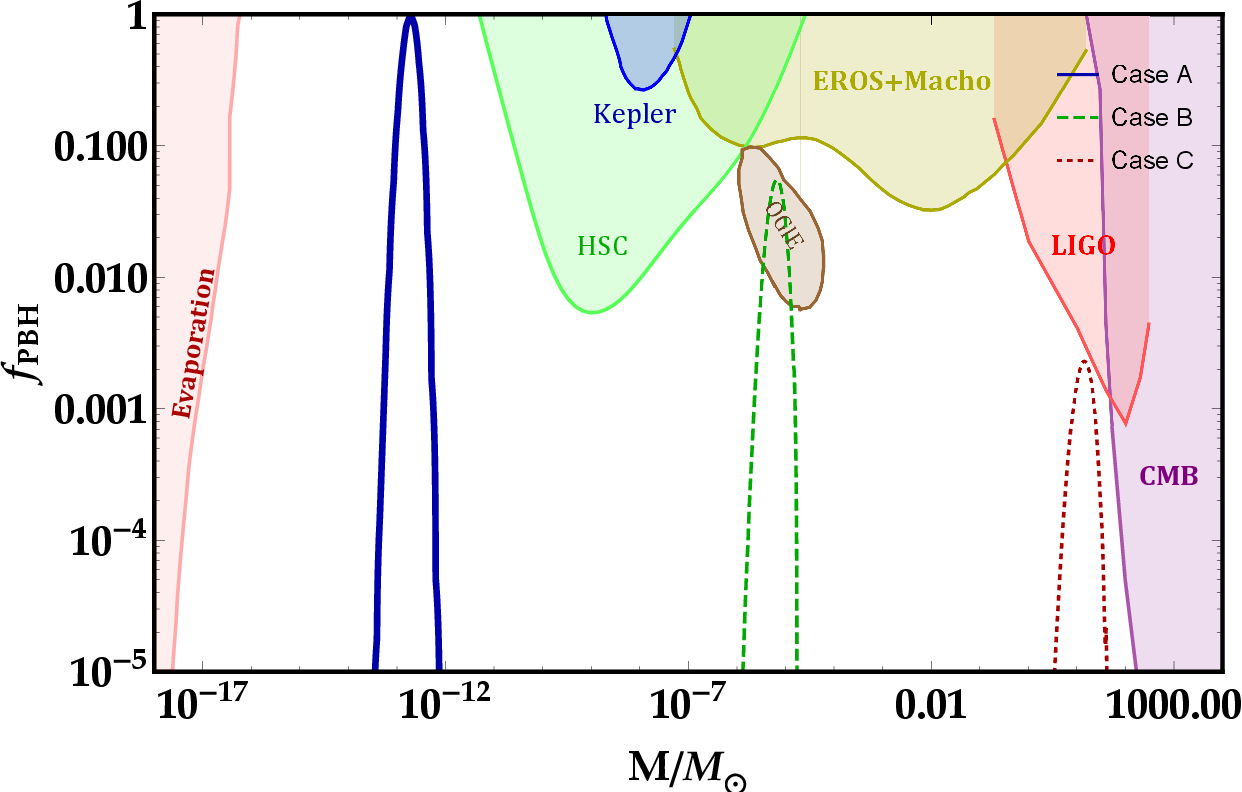}
	\caption{The abundances of PBHs with respect to their masses $M$ for the parameter sets listed in Table~\ref{tab1}.
		The cases A, B, and C are respectively demonstrated by blue, green, and red curves.
	The colored areas demonstrate the latest observational restrictions on the PBHs  abundance such as: the restriction on CMB from the signature of the spherical accretion of PBHs inside halos (purple area) \cite{CMB}, the upper bound on  PBHs abundance derived from the LIGO-Virgo event consolidation rate \cite{Abbott:2019,Chen:2022,Boehm:2021,Kavanagh:2018} (the border of the red area), the permitted region for PBHs mass spectrum from the ultrashort-timescale microlensing events in the OGLE data \cite{OGLE} (brown area), the constraints of microlensing events from collaborations between MACHO \cite{Alcock:2001}, EROS \cite{EORS} (light brown area), Kepler \cite{Kepler}(blue area), Icarus \cite{Icarus}, and Subaru-HSC \cite{subaro} (green area), the   evaporation constraints consist of  extragalactic $\gamma$-ray background \cite{EGG}, galactic center 511 keV $\gamma$-ray line (INTEGRAL) \citep{Laha:2019}, and effects on CMB spectrum \cite{Clark} (pink area).}
	\label{FPBHsSUSY}
\end{figure}

	%%%%%%%%%%%%%%%%%%%%%%%%%%%%%%%%%%%%%%%%%%
The objective of this section is to estimate the
PBHs abundance.
 As previously mentioned,  upon the re-entry of the enhanced scales to the horizon in the RD era, overdense zones start to collapse to form PBHs.
The mass of PBHs is contingent upon the horizon mass and determined by
\ba
\label{Mpbheq}
M(k) \simeq M_{\odot} \left(\frac{\gamma}{0.2} \right) \left(\frac{10.75}{g_{*}} \right)^{1/6} \left(\frac{k}{1.9\times 10^{6}\rm Mpc^{-1}} \right)^{-2},
\ea
where, $\gamma= (\frac{1}{\sqrt{3}})^{3}$ \cite{carr:1975} denotes the collapse efficiency, $g_{*}=106.75$ signifies the effective number of relativistic species at thermalization during the RD era.
In the following, employing the Press-Schechter theory and considering the Gaussian statistics for the distribution of curvature perturbations, the production rate of PBHs can be computed as follows~\cite{Tada:2019,young:2014}
\be
\label{betta}
\beta(M)=\int^{\infty}_{\delta_{c}}\frac{{\rm d}\delta}{\sqrt{2\pi\sigma^{2}(M)}}\exp\left(\frac{-\delta^{2}}{2\sigma^{2}(M)}\right).
\ee
Here, $\delta_{c}$ denotes the threshold value of density perturbations required for PBHs production. In this analysis, we adopt $\delta_{c}=0.4$ based on previous studies \cite{Musco:2013,Harada:2013}.
Additionally, the coarse-grained density contrast $\sigma^{2}(M)$, with the smoothing scale $k$, is defined as follows
\be
\label{Sigma}
\sigma_{k}^{2}=\left(\frac{2}{3} \right)^{4} \int^{\infty}_{0} \frac{{\rm d}q}{q} W^{2}(q/k)(q/k)^{4} {\cal P}_{s}(q),
\ee
where ${\cal P}_{s}$ is the scalar power spectrum  and  $W(x)=\exp{\left(-x^{2}/2 \right)}$ represents the Gaussian window function.
Finally, the PBHs abundance could be estimated as follows
\be
\label{fPBH}
f_{\rm{PBH}}(M)\simeq \frac{\Omega_{\rm {PBH}}}{\Omega_{\rm{DM}}}= \frac{\beta(M)}{1.84\times10^{-8}}\left(\frac{\gamma}{0.2}\right)^{3/2}\left(\frac{g_*}{10.75}\right)^{-1/4}
\left(\frac{0.12}{\Omega_{\rm{DM}}h^2}\right)
\left(\frac{M}{M_{\odot}}\right)^{-1/2},
\ee
where $M$ is given by Eq.~(\ref{Mpbheq}) and $\Omega_{\rm DM}h^2=0.12$ signifies the present density parameter of dark matter defined by Planck 2018 data \cite{akrami:2020}.
Utilizing ${\cal P}_s$ from solving the MS Eq.~(\ref{M.S}) and using Eqs. (\ref{Mpbheq})-(\ref{fPBH}),
the abundance of PBHs can be calculated. The obtained results for all cases in this study are reported in Table \ref{tab2} as well as Fig.~\ref{FPBHsSUSY}.

As illustrated in Fig. \ref{FPBHsSUSY}, Case A corresponding to
PBHs with mass $10^{-13}M_{\odot}$, roughly constitute
$99\%$ of the total dark matter content of the universe.
In Case B, the abundance of PBHs is $f_{\text{PBH}}^{\text{peak}} = 0.0546$ within the mass range of $10^{-6}M_{\odot}$.
Note that the results of this case is completely consistent with the allowed domain of the OGLE data.
Moreover, PBHs in Case C are generated with masses about $14.38M_{\odot}$, features $f_{\text{PBH}}^{\text{peak}} = 0.0023$.
As depicted in Fig.~\ref{FPBHsSUSY}, the LIGO-Virgo measurements place an upper limit on the abundance of this category of PBHs.
In addition, the abundance of PBHs for Case D, $f_{\text{PBH}}^{\text{peak}} \simeq 0$, reveals that it has inconsiderable contribution to dark matter.
However, the spectrum of the produced GWs from this case is consistent with the NANOGrav 15-year data.
\section{Investigation of Secondary GWs}
\label{sec5}
\begin{figure}[H]
		\centering
\hspace*{-1cm}
\includegraphics[scale=0.55]{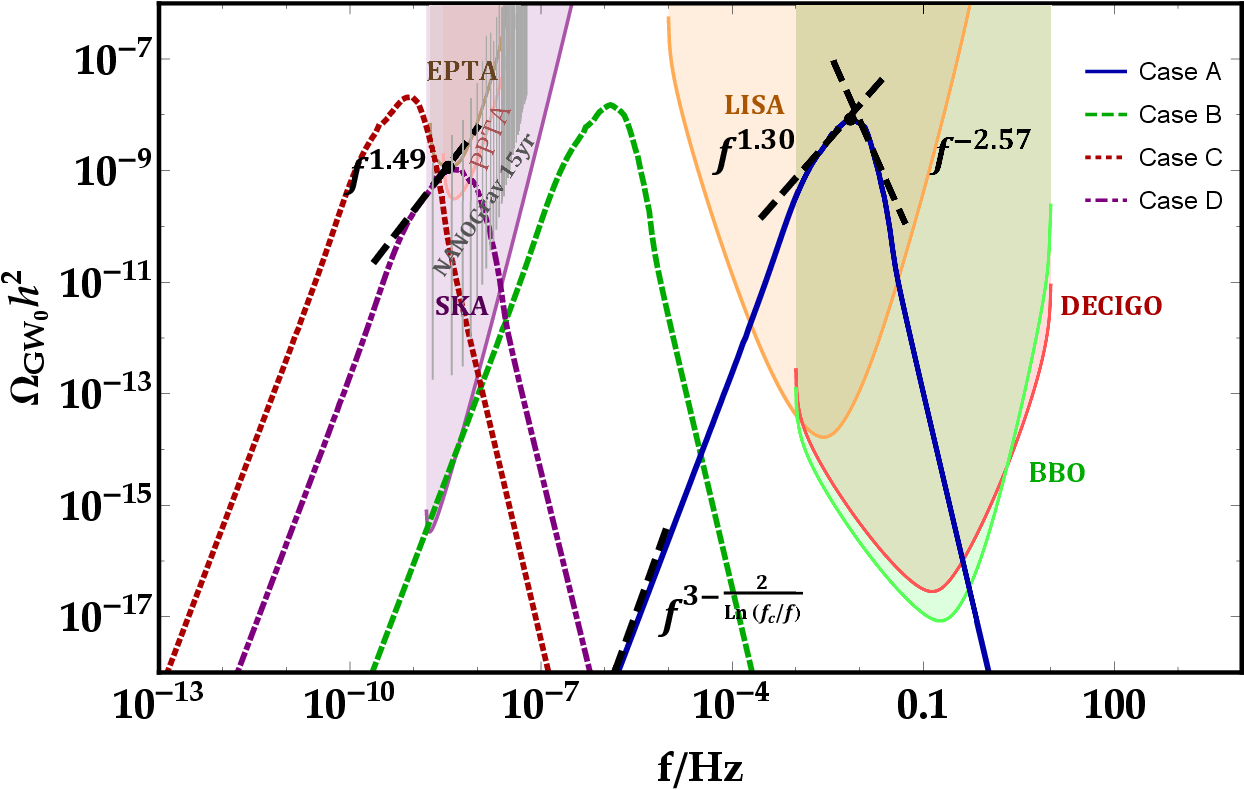}
\caption{The spectra of the present energy density parameter of the secondary gravitational waves $\Omega_{\rm GW_0}$
	in terms of frequency relevant to the
	low-scale SB SUSY potential in the GB framework for parameter sets of Table \ref{tab1}.
	The cases A, B, C, and D are respectively indicated by blue, green, red and purple curves.
	The precision of our forecasts can be validated through the data of different detectors like
	the European PTA (EPTA) \cite{EPTA-a,EPTA-b,EPTA-c,EPTA-d}, the Square Kilometer Array (SKA) \cite{ska,skaCarilli:2004,skaWeltman:2020}, Laser Interferometer Space Antenna (LISA) \cite{ligo-a,ligo-b,lisa,lisa-a}, BBO observatories \cite{Yagi:2011BBODECIGO,Yagi:2017BBODECIGO,Harry:2006BBO,Crowder:2005BBO,Corbin:2006BBO}, DECIGO \cite{Yagi:2011BBODECIGO,Yagi:2017BBODECIGO,Seto:2001DECIGO,Kawamura:2006DECIGO,Kawamura:2011DECIGO}, and the NANOGrav 15-year data \cite{Agazie1:2023,Agazie2:2023,Agazie3:2023,Agazie4:2023}.
		}\label{OMEGA}
	\end{figure}

Gravitational waves are produced in the RD era, when the enhanced scales of the perturbed modes re-enter the horizon.
Different detectors in various sensibility domains are able to identify these GWs \cite{Inomata:2019-a,Matarrese:1998,Mollerach:2004,Saito:2009,Garcia:2017,Cai:2019-a,Cai:2019-b,Cai:2019-c,Bartolo:2019-a,Bartolo:2019-b,Wang:2019,Fumagalli:2020b,Domenech:2020a,Domenech:2020b,Hajkarim:2019,Kohri:2018,Xu:2020,Fu:2020}.
In this discussion, we examine the generation of secondary GWs within the GB framework, incorporating low-scale SB SUSY potential.
Upon the end of the inflationary phase,
gradual decaying of the inflaton, results in  thermalizing the universe to join the RD era.
Consequently, the inflaton field has a negligible impact on cosmic evolution during the RD epoch. Therefore, the standard Einstein formulation can be employed to investigate the production of secondary GWs during this era.
Therefore, during the RD era, the energy density parameter of GWs can be calculated as follows \cite{Ananda:2007,Baumann:2007,Kohri:2018,Lu:2019}
\be
	\label{OGW}
	\Omega_{\rm GW}(k,\eta)=\frac{1}{6}\left(\frac{k}{aH}\right)^{2}\int_{0}^{\infty}dv\int_{|1-v|}^{|1+v|}du\left(\frac{4v^{2}-\left(1-u^{2}+v^{2}\right)^{2}}{4uv}\right)^{2}\overline{I_{RD}^{2}(u,v,x)}\mathcal{P}_{s}(ku)\mathcal{P}_{s}(kv),
\ee
where $\eta$ represents the conformal time, and the time-averaged value of the source terms is expressed as follows
\ba
	\overline{I_{\rm RD}^{2}(u,v,x\to\infty)}= & \frac{1}{2x^{2}}\Bigg[\left(\frac{3\pi\left(u^{2}+v^{2}-3\right)^{2}\Theta\left(u+v-\sqrt{3}\right)}{4u^{3}v^{3}}+\frac{T_{c}(u,v,1)}{9}\right)^{2}& +\left(\frac{\tilde{T}_{s}(u,v,1)}{9}\right)^{2}\Bigg].
	\label{IRD2b}
\ea
Here, $\Theta$ denotes the Heaviside theta function, which is defined as $\Theta(x\geq 0)=1$ and $\Theta(x<0)=0$.
Additionally, the functions utilized in Eq.~(\ref{IRD2b}) are defined as follows
\begin{align}
	T_{c}= & -\frac{27}{8u^{3}v^{3}x^{4}}\Bigg\{-48uvx^{2}\cos\left(\frac{ux}{\sqrt{3}}\right)\cos\left(\frac{vx}{\sqrt{3}}\right)\left(3\sin(x)+x\cos(x)\right)+
	\nonumber\\
	& 48\sqrt{3}x^{2}\cos(x)\left(v\sin\left(\frac{ux}{\sqrt{3}}\right)\cos\left(\frac{vx}{\sqrt{3}}\right)+u\cos\left(\frac{ux}{\sqrt{3}}\right)\sin\left(\frac{vx}{\sqrt{3}}\right)\right)+
	\nonumber\\
	& 8\sqrt{3}x\sin(x)\Bigg[v\left(18-x^{2}\left(u^{2}-v^{2}+3\right)\right)\sin\left(\frac{ux}{\sqrt{3}}\right)\cos\left(\frac{vx}{\sqrt{3}}\right)+
	\nonumber\\
	& u\left(18-x^{2}\left(-u^{2}+v^{2}+3\right)\right)\cos\left(\frac{ux}{\sqrt{3}}\right)\sin\left(\frac{vx}{\sqrt{3}}\right)\Bigg]+
	\nonumber\\
	& 24x\cos(x)\left(x^{2}\left(-u^{2}-v^{2}+3\right)-6\right)\sin\left(\frac{ux}{\sqrt{3}}\right)\sin\left(\frac{vx}{\sqrt{3}}\right)+
	\nonumber\\
	& 24\sin(x)\left(x^{2}\left(u^{2}+v^{2}+3\right)-18\right)\sin\left(\frac{ux}{\sqrt{3}}\right)\sin\left(\frac{vx}{\sqrt{3}}\right)\Bigg\}
	\nonumber\\
	& -\frac{\left(27\left(u^{2}+v^{2}-3\right)^{2}\right)}{4u^{3}v^{3}}\Bigg\{\text{Si}\left[\left(\frac{u-v}{\sqrt{3}}+1\right)x\right]-\text{Si}\left[\left(\frac{u+v}{\sqrt{3}}+1\right)x\right]
	\nonumber\\
	& +\text{Si}\left[\left(1-\frac{u-v}{\sqrt{3}}\right)x\right]-\text{Si}\left[\left(1-\frac{u+v}{\sqrt{3}}\right)x\right]\Bigg\},
	\label{Tc}
\end{align}

\begin{align}
	T_{s}=& \frac{27}{8u^{3}v^{3}x^{4}}\Bigg\{48uvx^{2}\cos\left(\frac{ux}{\sqrt{3}}\right)\cos\left(\frac{vx}{\sqrt{3}}\right)\left(x\sin(x)-3\cos(x)\right)-
	\nonumber\\
	& 48\sqrt{3}x^{2}\sin(x)\left(v\sin\left(\frac{ux}{\sqrt{3}}\right)\cos\left(\frac{vx}{\sqrt{3}}\right)+u\cos\left(\frac{ux}{\sqrt{3}}\right)\sin\left(\frac{vx}{\sqrt{3}}\right)\right)+
	\nonumber\\
	& 8\sqrt{3}x\cos(x)\Bigg[v\left(18-x^{2}\left(u^{2}-v^{2}+3\right)\right)\sin\left(\frac{ux}{\sqrt{3}}\right)\cos\left(\frac{vx}{\sqrt{3}}\right)+
	\nonumber\\
	& u\left(18-x^{2}\left(-u^{2}+v^{2}+3\right)\right)\cos\left(\frac{ux}{\sqrt{3}}\right)\sin\left(\frac{vx}{\sqrt{3}}\right)\Bigg]+
	\nonumber\\
	& 24x\sin(x)\left(6-x^{2}\left(-u^{2}-v^{2}+3\right)\right)\sin\left(\frac{ux}{\sqrt{3}}\right)\sin\left(\frac{vx}{\sqrt{3}}\right)+
	\nonumber\\
	& 24\cos(x)\left(x^{2}\left(u^{2}+v^{2}+3\right)-18\right)\sin\left(\frac{ux}{\sqrt{3}}\right)\sin\left(\frac{vx}{\sqrt{3}}\right)\Bigg\}-\frac{27\left(u^{2}+v^{2}-3\right)}{u^{2}v^{2}}+
	\nonumber\\
	& \frac{\left(27\left(u^{2}+v^{2}-3\right)^{2}\right)}{4u^{3}v^{3}}\Bigg\{-\text{Ci}\left[\left|1-\frac{u+v}{\sqrt{3}}\right|x\right]+\ln\left|\frac{3-(u+v)^{2}}{3-(u-v)^{2}}\right|+
	\nonumber\\
	& \text{Ci}\left[\left(\frac{u-v}{\sqrt{3}}+1\right)x\right]-\text{Ci}\left[\left(\frac{u+v}{\sqrt{3}}+1\right)x\right]+\text{Ci}\left[\left(1-\frac{u-v}{\sqrt{3}}\right)x\right]\Bigg\}.
	\label{Ts}
\end{align}
Moreover, the sine-integral $\text{Si}(x)$ and cosine-integral $\text{Ci}(x)$ functions are defined as
\be
	\label{SiCi}
	\text{Si}(x) \equiv \int_{0}^{x}\frac{\sin(y)}{y}dy,\qquad\text{Ci}(x) \equiv -\int_{x}^{\infty}\frac{\cos(y)}{y}dy.
\ee
The function $\tilde{T}_{s}(u,v,1)$ from Eq.~\eqref{IRD2b}, is expressed as
\be
	\label{Tst}
	\tilde{T}_{s}(u,v,1)=T_{s}(u,v,1)+\frac{27\left(u^{2}+v^{2}-3\right)}{u^{2}v^{2}}-\frac{27\left(u^{2}+v^{2}-3\right)^{2}}{4u^{3}v^{3}}\ln\left|\frac{3-(u+v)^{2}}{3-(u-v)^{2}}\right|.
\ee
The current energy spectrum of the GWs is defined as follows \cite{Inomata:2019-a}
\ba
	\label{OGW0}
	\Omega_{\rm GW_0}h^2 = 0.83\left( \frac{g_{*}}{10.75} \right)^{-1/3}\Omega_{\rm r_0}h^2\Omega_{\rm{GW}}(\eta_c,k)\,,
\ea
wherein, $\Omega_{\rm{r_0}}h^2\simeq 4.2\times 10^{-5}$ represents the radiation density parameter at the current epoch, and $g_*\simeq106.75$ denotes the effective degrees of freedom in the energy density at $\eta_c$. Also, the frequency in terms of the wavenumber reads
\ba
	\label{k_to_f}
	f=1.546 \times 10^{-15}\left( \frac{k}{{\rm Mpc}^{-1}}\right){\rm Hz}.
\ea
Finally, we can evaluate the current energy density parameter of generated GWs from PBH formation in our model by utilizing the power spectrum derived from the MS Eq.~(\ref{M.S}) and using Eqs.~(\ref{OGW})-(\ref{k_to_f}).
Fig. \ref{OMEGA} indicates our obtained results for  $\Omega_{\rm GW_0}$ accompanied by the sensibility domains of GWs detectors. It can be inferrable from this figure that,
For cases A, B, C, and D, the spectra of $\Omega_{\rm GW_0}$ are located at different frequencies.
About Case A, associated with asteroid-mass PBHs, the peak of $\Omega_{\rm GW_0}$ spectrum is observed in the mHz frequency ranges, detectable by observatories such as LISA \cite{ligo-a,ligo-b,lisa,lisa-a}, BBO \cite{Yagi:2011BBODECIGO,Yagi:2017BBODECIGO,Harry:2006BBO,Crowder:2005BBO,Corbin:2006BBO}, and DECIGO \cite{Yagi:2017BBODECIGO,Seto:2001DECIGO,Kawamura:2006DECIGO,Kawamura:2011DECIGO}.
The spectra of $\Omega_{\rm GW_0}$ for Cases B, C and D, are located in  the detection area of the SKA detector \cite{ska,skaCarilli:2004,skaWeltman:2020}. Note that the spectrum of $\Omega_{\rm GW_0}$ for Case D
crosses the sensitivity areas of the NANOGrav and PPTA observatories as well (See Fig.~\ref{OMEGA}).
Several GWs observatories can test the estimated $\Omega_{\rm GW_0}$ spectrum for all cases considered in this study. This means that the data of several observatories could be used to evaluate the credibility  of this model.
Lastly, we scrutinize the
evolution
of the $\Omega_{\rm GW_0}$
spectrum within various  frequency zones.
It has been verified that in proximity to
the peak location, the $\Omega_{\rm GW_0}$ behaves as a power-law function in terms of the frequency, i.e., $\Omega_{\rm GW_0} (f) \sim f^{n}$
\cite{Fu:2019vqc,Fu:2020lob,Bagui:2021dqi,Xu,Kuroyanagi}.
Table \ref{tableGWs0} shows the frequencies and height of the peaks for all cases.
Additionally, our findings consistent with the analytic result, $\Omega_{\rm GW_0}\sim f^{3-2/\ln\left(f_{c}/f\right)}$ in the infrared regime $f \ll f_c$, derived by \cite{Yuan:2020,Sasaki:2020}, as depicted in Fig. \ref{OMEGA}.
 \begin{table}[H]
	\centering
	\caption{The frequency and height related to the peak of the $\Omega_{\rm GW_0}$ spectrum and power index $n$ in different frequency reigns $f\ll f_{c}$, $f<f_{c}$ and $f>f_{c}$ for each case of the model.}
	\scalebox{1}[1] {
		\begin{tabular}{cccccc}
			\hline
			\#  & $\qquad\qquad$ $f_{c}$ $\qquad\qquad$ & $\quad$ $\Omega_{\rm GW_0}\left(f_{c}\right)$ $\quad$ & $\quad$ $n_{f\ll f_{c}}$ $\quad$ & $\quad$ $n_{f<f_{c}}$ $\quad$ & $\quad$ $n_{f>f_{c}}$\tabularnewline
			\hline
			\hline
			$\rm{Case~}A$ & $7.353\times10^{-3}$ & $8.441\times10^{-9}$ & $3.0$ & $1.30$ & $-2.57$
			\tabularnewline
			\hline
			$\rm{Case~}B$ & $1.231\times10^{-6}$ & $1.507\times10^{-8}$ & $2.92$ & $1.33$ & $-1.87$ \tabularnewline
			\hline
			$\rm{Case~}C$ & $8.354\times10^{-10}$ & $2.064\times10^{-8}$ & $2.82$ & $1.09$ & $-1.70$ \tabularnewline
			\hline
			$\rm{Case~}D$ & $3.379\times10^{-9}$ & $1.157\times10^{-9}$ & $2.84$ & $1.49$ & $-1.27$ \tabularnewline
			\hline
		\end{tabular}
	}
	\label{tableGWs0}
\end{table}
In our model, the detection of induced GW signals presents a unique opportunity to discern the presence of PBHs. Recently, the NANOGrav pulsar timing array team unveiled low-frequency stochastic gravitational wave background through 15 years of data collection \cite{Agazie1:2023,Agazie2:2023,Agazie3:2023,Agazie4:2023}. As mentioned, the spectrum of $\Omega_{\rm GW_0}$ for Case D
{crosses
the sensitivity ranges of the PPTA and NANOGrav
observatories
(refer to Fig. \ref{OMEGA}).
The low frequency  GWs signal in the nano-Hz domain can be characterized as $\Omega_{\rm{GW}}(f)\sim f^{5-\gamma}$, where $\gamma$ denotes the spectral index and its value is determined by the NANOGrav teamwork as $\gamma=3.2\pm 0.6$
\cite{BabichevFitNANOGrav:2023,VagnozziFitNANOGrav:2023,AntoniadisFitNANOGrav:2023a,AntoniadisFitNANOGrav:2023b,AntoniadisFitNANOGrav:2023c}.
Regarding the last row of Table \ref{tableGWs0} it is obvious that, the power index $n=1.49$ for Case D results in $\gamma=5-n=3.51$, which aligns with the observed findings.
%

%%%%%%%%%%%%%%%%%%%%%%%%%%%%%%%%%%%%%%%%%%%%%%%%%%
\section{Results and discussion}
\label{sec6}
%%%%%%%%%%%%%%%%%%%%%%%%%%%%%%%%%%%%%%%%%%%%%%%%%%
%
This study explored  PBHs formation within an inflationary framework involving a coupling between the scalar field and the GB term. Moreover, we examined the low-scale SB SUSY potential within this context. In the presence of the GB coupling term, the inflaton traverses near a non-trivial fixed point and undergoes an USR epoch. Consequently, the primordial curvature perturbations can experience enhancement during this era. %Here, we have investigated a potential that has been excluded
The low-scale SB SUSY potential in the standard framework of inflation is ruled out by the Planck 2018 observations \cite{akrami:2020}. This characteristic led us to reconsider it within the GB model to attain
a tenable model of inflation.
By employing a suitable GB coupling function and adjusting the free parameters of model
($M$, $\beta$, $\xi_{0}$, $\xi_{1}$, $\phi_{c}$, $V_0$, $\alpha$), we effectively aligned the SB SUSY potential with the findings of CMB measurements. In addition, during the USR phase, the amplitude of perturbations can increase, leading to the production of PBHs. Utilizing the precise solution of the background equations, we illustrated the evolution of the
scalar field $\phi$, the Hubble parameter $H$, the first and second slow-roll parameters $(\epsilon_1$ and $\epsilon_2$)
, $c_s^{2}$, and $c_t^{2}$ in Fig. \ref{Fig1} as functions of the number of $e$-folds. It is evident from this figure that during the USR phase, the slow-roll conditions are upheld by the first parameter ($\epsilon_1 \ll 1$) and violated by the second parameter ($\left|\epsilon_2\right| \geq 1$).
In this regard,
the numerical solutions of the MS Equation (\ref{M.S}) were derived to obtain the exact curvature power spectra for cases of Table \ref{tab1}.
Furthermore, Figure \ref{Fig2} illustrates that, the determined power spectra contain peaks with sufficient amplitudes to form PBHs on small scales. Also, the value of the power spectrum at large scales is consistent with the Planck 2018 observations.
Our calculations yield $n_s=0.970$, that aligns with $68\,\%\,\rm CL$ of the Planck 2018 TT, TE, EE + lowE + lensing + BK18 + BAO \cite{akrami:2020,Campeti:2022}, and $r=0.005$, which aligns with the BICEP/Keck 2018 data ($r<0.036$ at $95\,\%\,\rm CL$) \cite{akrami:2020,Ade:2021}.
Hence, through the utilization of the GB framework the SB SUSY potential is revitalized.
Moreover, this choice enables us to generate PBHs with suitable abundance and GWs within the sensitivity range of observatories. Also, our analysis confirms that our model satisfies the swampland criteria.
In Fig.~\ref{FPBHsSUSY}, Case A corresponding to PBHs with masses around $10^{-13}M_{\odot}$, constituting nearly all of the dark matter. In Case B, PBH abundance peaks at $f_{\text{PBH}}^{\text{peak}} = 0.0546$ with mass around $10^{-6}M_{\odot}$, aligning with observed microlensing events in the OGLE data. Case C features PBHs with a mass of approximately $14.38M_{\odot}$ and a peak abundance $f_{\text{PBH}}^{\text{peak}} = 0.0023$, constrained by upper limits from
LIGO-Virgo data.
The PBHs abundance of Case D is
approximately $f_{\text{PBH}}^{\text{peak}} \simeq 0$, indicating a negligible contribution of dark matter. However, the produced spectrum of GWs from this case aligns with the NANOGrav 15-year data.
Lastly, we examined the production of induced GWs subsequent to the formation of PBHs for all cases of our model.
In Case A, linked to asteroid-mass PBHs, the peak of $\Omega_{\rm GW_0}$ spectrum occurs in the mHz frequency range,
which can be detected
by LISA, BBO, and DECIGO.
The spectra of $\Omega_{\rm GW_0}$ for Cases B and C, are located near the $f_{c} \sim 10^{-6}$ Hz and $f_{c} \sim 10^{-10}$ Hz, and can be tested by the SKA observatory.
Additionally, the spectrum of $\Omega_{\rm GW_0}$ for Case D intersects the sensitivity ranges of the PPTA and NANOGrav detectors (refer to Fig. \ref{OMEGA}).
Consequently, the validity of our outcomes could be assessed by analyzing the obtained data of these observatories.
In the final phase, we calculated the power-law behaviour of $\Omega_{\rm GW_0}$ spectra ($\Omega_{\rm GW_0}(f) \sim f^n$) for all Cases across different frequency ranges. Table \ref{tableGWs0} presents the peak frequencies of $\Omega_{\rm GW_0}$ spectra and the estimated values for the power indices, $n$, for all cases in three frequency regions $f\ll f_{c}$, $f < f_c$, and $f > f_c$. It can be deduced that the resulting values for $n$ in the infrared region $f \ll f_c$ could align with the logarithmic equation $n=3-2/\ln(f_c/f)$. In Case D,  our
estimation, $\gamma=5-n=3.51$, aligns with the observed results from NANOGrav 15-year data. It is worth to mention that, in the present work, we could rescue the low-scale SB SUSY potential Eq.~(\ref{eqv}) in the GB framework and succeeded in generating PBHs and SGWs in this framework, with the help of coupling function Eqs.~(\ref{eqxi}) and (\ref{eqxi12}). However, to the best of our knowledge, with the current observational data there are no distinguishable features between different inflationary models to yield PBHs and SGWs. So, the hope is that, the upcoming observational data could provide suitable feature to distinguish between PBHs models.
%
%--------------------------------------------------------------------------------------------------------------------
%========================================Acknowledgements=====================================================
\section{Acknowledgements}\label{sec7}
The authors thank the referee for his/her valuable comments.

%%%%%%%%%%%%%%%%%%%%%%%%%%%%%%%%%%%%%%%%%%%%%%%%%%%%%%%%%%%%%%%%%%%	
	%========================References============================
	%\subsection*{References}
	

\begin{thebibliography}{0}
\expandafter\ifx\csname natexlab\endcsname\relax\def\natexlab#1{#1}\fi
\expandafter\ifx\csname bibnamefont\endcsname\relax
  \def\bibnamefont#1{#1}\fi
\expandafter\ifx\csname bibfnamefont\endcsname\relax
  \def\bibfnamefont#1{#1}\fi
\expandafter\ifx\csname citenamefont\endcsname\relax
  \def\citenamefont#1{#1}\fi
\expandafter\ifx\csname url\endcsname\relax
  \def\url#1{\texttt{#1}}\fi
\expandafter\ifx\csname urlprefix\endcsname\relax\def\urlprefix{URL }\fi
\providecommand{\bibinfo}[2]{#2}
\providecommand{\eprint}[2][]{\url{#2}}

\end{thebibliography}


\begin{thebibliography}{100}

\bibitem{zeldovich:1967}
Ya. B. Zel'dovich, and I. D. Novikov, Sov. Astron. {\bf 10}, 602 (1967).
%
\bibitem{Hawking:1971}
S. Hawking, Mon. Not. R. Astron. Soc. {\bf 152}, 75 (1971).
%
\bibitem{Hawking:1974}
S. Hawking, Nature {\bf 248}, 30 (1974).
%%3 Proposal Ali
%
\bibitem{Hawking:1976}
S. Hawking, Commun. Math. Phys. {\bf 46}, 206 (1976).
%%[Erratum:Commun Math. Phys. 43, 199 (1975).]
%%48 Rezazadeh
%
\bibitem{carr:1975}
B. J. Carr, Astrophys. J. {\bf 201}, 1 (1975).
%5 Proposal Ali
%
\bibitem{carr:1974}
B. J. Carr, and S.W. Hawking, Mon. Not. R. Astron. Soc. {\bf 168}, 399 (1974).	
%
\bibitem{Page:1976}
D. N. Page, and S. Hawking, Astrophys. J. {\bf 206}, 1 (1976).
%4 Proposal Ali	
%
\bibitem{Olmo:2011}
G. J. Olmo, Int. J. Mod. Phys. D {\bf 20} (2011).
%9 Odintsov		
%
\bibitem{Cruz-Dombriz:2012}
A. de la Cruz-Dombriz, and D. Saez-Gomez, Entropy {\bf 14}, 1717 (2012).
%8 Odintsov		
%
\bibitem{Chapline:1975}
G. F. Chapline, Nature {\bf 253}, 251 (1975).
%5 kawai Nature Londen
%
\bibitem{Polnarev:1985}
A. G. Polnarev, and M. Y. Khlopov, Sov. Phys. Usp. {\bf 28}, 213 (1985).
%4 kawai		
%
\bibitem{Ivanov:1994}
P. Ivanov, P. Nasselsky, and I. D. Novikov, Phys. Rev. D {\bf 50}, 7173 (1994).
%

%NEW NEW Refs
\bibitem{Qiu:2023NEWQiu}
T. Qiu, W. Wang, and R. Zheng, Phys. Rev. D {\bf 107}, 083018 (2023).
%Generation of primordial black holes from an inflation model with modified dispersion relation
%Taotao Qiu, Wenyi Wang, and Ruifeng Zheng
%Phys. Rev. D 107, 083018 – Published 11 April 2023

\bibitem{SChoudhury:2014}
S. Choudhury, and A. Mazumdar, Phys. Lett. B {\bf 733}, 270 (2014).
%
\bibitem{Solbi-a:2021}
M. Solbi, and K. Karami,  Eur. Phys. J. C {\bf81}, 884 (2021).
%
\bibitem{Solbi-b:2021}
M. Solbi, and K. Karami, J. Cosmol. Astropart. Phys. {\bf08}, 056  (2021).
%
\bibitem{Teimoori-b:2021}
Z. Teimoori, K. Rezazadeh, M. A. Rasheed, and K. Karami, J. Cosmol. Astropart. Phys. {\bf10}, 018 (2021).
%
\bibitem{Rezazadeh:2021}
K. Rezazadeh, Z. Teimoori, and K. Karami,  Eur. Phys. J. C {\bf82}, 758 (2022).
%
\bibitem{YCai:2021}
Y. Cai, and Y. S. Piao, Phys. Rev. D {\bf 103}, 083521 (2021).
%
\bibitem{AChakraborty:2022}
A. Chakraborty, P. K. Chanda, K. L. Pandey, and S. Das, Astrophys. J. {\bf 932}, 119 (2022).
%
\bibitem{TPapanikolaou:2022}
T. Papanikolaou, C. Tzerefos, S. Basilakos, and E. N. Saridakis, J. Cosmol. Astropart. Phys. {\bf 2022}, 13 (2022).
%		
\bibitem{TPapanikolaou:2023a}
T. Papanikolaou, C. Tzerefos, S. Basilakos, and E. N. Saridakis, Eur. Phys. J. C {\bf 83}, 31 (2023).
%
\bibitem{TPapanikolaou:2023b}
T. Papanikolaou, A. Lymperis, S. Lolab, and E. N. Saridakis, J. Cosmol. Astropart. Phys. {\bf 03}, 003 (2023).
%
\bibitem{GDomenech:2023}
G. Dom{\`e}nech, G. Vargas, and T. Vargas, J. Cosmol. Astropart. Phys. {\bf 03}, 002 (2024).
%
\bibitem{YCai:2023}
Y. Cai, M. Zhu, and Y. S. Piao, Phys. Rev. Lett. {\bf 133}, 021001 (2024).
%Y. Cai, M. Zhu, and Y. S. Piao, arXiv:2305.10933.	
		
		
		
		
\bibitem{Abbott:2016}
B. P. Abbott et al., (LIGO Scientific and Virgo Collaborations) Phys. Rev. D {\bf 93}, 122004 (2016).
%7 Proposal Ali      OK
%
\bibitem{Abbott:2016-a}
B. P. Abbott et al., (LIGO Scientific Collaboration and Virgo Collaboration), Phys. Rev. Lett.
{\bf 116}, 061102 (2016).
%
\bibitem{Abbott:2016-b}
B. P. Abbott et al., (LIGO Scientific Collaboration and Virgo Collaboration), Phys. Rev. Lett.
{\bf 116}, 241103 (2016).
%
\bibitem{Abbott:2017}
B. P. Abbott et al., Astrophys. J. Lett. {\bf 848}, L12 (2017).
%1 Odintsov
\bibitem{Abbott:2017-a}
B. P. Abbott et al., (LIGO Scientific Collaboration and Virgo Collaboration), Phys. Rev. Lett.
{\bf 116}, 221101 (2017).
%
\bibitem{Abbott:2017-b}
B. P. Abbott et al., (LIGO Scientific Collaboration and Virgo Collaboration), Astrophys. J.
{\bf 851}, L35 (2017).
%
\bibitem{Abbott:2017-c}
B. P. Abbott et al., (LIGO Scientific Collaboration and Virgo Collaboration), Phys. Rev. Lett.
{\bf 119}, 141101 (2017).
%
\bibitem{Abbott:2019}
B. P. Abbott, et al., (LIGO Scientific Collaboration and Virgo Collaboration), Phys. Rev. Lett. {\bf 123}, 161102 (2019).
%Abbott, Benjamin P and Abbott, R and Abbott, TD and Abraham, S and Acernese, Fausto and Ackley, K and Adams, C and Adhikari, Rana X and Adya, VB and Affeldt, C and others
%
\bibitem{OGLE}
H. Niikura et al., Phys. Rev. D \textbf{99}, 083503 (2019).
%
\bibitem{Alcock:2001}
C. Alcock et al., Astrophys. J. Lett. {\bf 550}, L169 (2001).
%
\bibitem{Gould:1992}
A. Gould, Astrophys. J. {\bf 386}, L5 (1992).
%49 Rezazadeh
%
\bibitem{Heydari:2021}
S. Heydari, and K. Karami, Eur. Phys. J. C {\bf82}, 83 (2022).		
%57 S
\bibitem{Heydari:2023b}
S. Heydari, and K. Karami, 	Eur. Phys. J. C {\bf 84}, 127 (2024).
%
\bibitem{Heydari:2023a}
S. Heydari, and K. Karami, J. Cosmol. Astropart. Phys. {\bf 02}, 047 (2024).
%
%\bibitem{Heydari:2023}
%S. Heydari, and K. Karami, arXiv:2309.01239.
%
\bibitem{Dalcanton:1994}
J. J. Dalcanton et al., Astrophys. J. {\bf 424}, 550 (1994).
%50 Rezazadeh
%
\bibitem{Sato-Polito:2019}
G. Sato-Polito, E. D. Kovetz, and M. Kamionkowski, Phys. Rev. D {\bf 100}, 063521 (2019).
%60 Rezazadeh
%
\bibitem{Jacobs:2015}
D. M. Jacobs et al., Mon. Not. Roy. Astron. Soc. {\bf 450}, 3418 (2015).
%57 Rezazadeh
%
\bibitem{Ali-Haimoud:2017}
Y. Ali-Haimoud, and M. Kamionkowski, Phys. Rev. D {\bf 95}, 043534 (2017).
%
\bibitem{Wang:2018}
S. Wang, Y. F. Wang, Q. G. Huang, and T. G. F. Li, Phys. Rev. Lett. {\bf 120}, 191102 (2018).
%
\bibitem{Laha:2020}
R. Laha, J. B. Munoz, and T. R. Slatyer, Phys. Rev. D {\bf 101}, 123514 (2020).
%
\bibitem{Teimoori:2021}
Z. Teimoori, K. Rezazadeh, and K. Karami, Astrophys. J. {\bf 915}, 118 (2021).
%
\bibitem{Heydari:2024}
S. Heydari, and K. Karami, arXiv:2405.08563.
%
\bibitem{Heydari:2022}
S. Heydari, and K. Karami, J. Cosmol. Astropart. Phys. {\bf 03}, 033 (2022).
%
\bibitem{Solbi:2024}
M. Solbi, and K. Karami, arXiv:2403.00021.
%arXiv:2403.00021
%
\bibitem{akrami:2020}
Y. Akrami et al., (Planck Collaboration), Astron. Astrophys. {\bf 641}, A10 (2020).
%
\bibitem{Dalianis:2019}
I. Dalianis, A. Kehagias, and G. Tringas, J. Cosmol. Astropart. Phys. {\bf 01}, 037 (2019).
%
\bibitem{mahbub:2020}
R. Mahbub, Phys. Rev. D {\bf 101}, 023533 (2020).
%
\bibitem{Kamenshchik:2019}
A. Y. Kamenshchik, A. Tronconi, T. Vardanyan, and G. Venturi, Phys. Lett. B {\bf 791}, 201
(2019).
%
\bibitem{Kawai:2021}
S. Kawai, and J. Kim, Phys. Rev. D \textbf{104}, 083545 (2021).
%26 Proposal Ali
%
\bibitem{Zhang:2022}
F. Zhang, Phys. Rev. D {\bf 105}, 063539 (2022).
%	
\bibitem{Kawaguchia:2023}
R. Kawaguchi, and Sh. Tsujikawa, Phys. Rev. D {\bf 107}, 063508 (2023).
%29 Proposal Ali
%
\bibitem{Ashrafzadeh:2024}
A. Ashrafzadeh, and K. Karami, Astrophys. J. {\bf 965}, 11 (2024).
%
\bibitem{Fu:2019}
C. Fu, P. Wu, and H. Yu, Phys. Rev. D {\bf 100}, 063532 (2019).
%
\bibitem{Mishra:2020}
S. S. Mishra, and V. Sahni, J. Cosmol. Astropart. Phys. {\bf 04},  007 (2020).
%
\bibitem{Villanueva-Domingo:2021}
P. Villanueva-Domingo, O. Mena, and S. Palomares-Ruiz, Front. Astron. Space. Sci. {\bf 8}, 103389 (2021).
%Villanueva-Domingo, P., Mena, O. & Palomares-Ruiz, S.
%
\bibitem{Gao:2021}
Q. Gao, Y. Gong, and Zh. Yi, Nucl. Phys. B {\bf 969}, 115480 (2021).
%27 Proposal Ali Gao, Q., Gong, Y., & Yi, Zh.,
%		
\bibitem{Lin:2020}
J. Lin et al., Phys. Rev. D {\bf 101}, 103515 (2020).
%24 Proposal Ali
%
\bibitem{SChoudhury:2023}
S. Choudhury, S. Panda, and M. Sami, Phys. Lett. B {\bf 845}, 138123 (2023).
%Choudhury S, Panda S, Sami M. PBH formation in EFT of single field inflation with sharp transition. Physics Letters B. 2023 Oct 10;845:138123.      FPBHs
%

%
\bibitem{Bhattacharya:2023}
S. Bhattacharya et al., Galaxies {\bf 11}, 35 (2023).
%
\bibitem{SMittal:2022}
S. Mittal, A. Ray, G. Kulkarni and B. Dasgupta, J. Cosmol. Astropart. Phys. {\bf 2022}, 30 (2022).
%Mittal, Shikhar and Ray, Anupam and Kulkarni, Girish and Dasgupta, Basudeb       21cm
%
\bibitem{GDomenech:2021a}
G. Dom{\`e}nech, Universe. {\bf 7}, 398 (2021).
%
\bibitem{GDomenech:2021b}
G. Dom{\`e}nech, Ch. Lin, and M. Sasaki, J. Cosmol. Astropart. Phys. {\bf 04}, 062 (2021).
%
\bibitem{GDomenech:2024}
G. Dom{\`e}nech, and M. Sasaki, Class. Quantum Grav. {\bf 41}, 143001 (2024).
%
\bibitem{MRGangopadhyay:2023b}
M. R. Gangopadhyay et al., arXiv:2309.03101.
%
\bibitem{GBhattacharya:2023}
G. Bhattacharya et al., arXiv:2309.00973.
%









\bibitem{Banerjee:2022NEWPapanikolaou}
S. Banerjee, T. Papanikolaou, and E. N. Saridakis, Phys. Rev. D {\bf 106}, 124012 (2022).
%PBHs
%Constraining F(R) bouncing cosmologies through primordial black holes

\bibitem{Papanikolaou:2024NEWPapanikolaou}
T. Papanikolaou, S. Banerjee, Y. F. Cai, S. Capozziello, and E. N. Saridakis, J. Cosmol. Astropart. Phys. {\bf 06}, 066 (2024).
%PBHs GWs
%Primordial black holes and induced gravitational waves in non-singular matter bouncing cosmology
%Published 26 June 2024 • © 2024 The Author(s)
%Journal of Cosmology and Astroparticle Physics, Volume 2024, June 2024
%Citation Theodoros Papanikolaou et al JCAP06(2024)066
%DOI 10.1088/1475-7516/2024/06/066







\bibitem{Basilakos:2024NEWPapanikolaou}
S. Basilakos, D. V. Nanopoulos, T. Papanikolaou, E. N. Saridakis, and C. Tzerefos, Phys. Lett. B {\bf 849}, 138446 (2024).
%GWs nHz
%Induced gravitational waves from flipped SU(5) superstring theory at nHz


\bibitem{Tzerefos:2023NEWPapanikolaou}
C. Tzerefos, T. Papanikolaou, E. N. Saridakis, and S. Basilakos, Phys. Rev. D {\bf 107}, 124019 (2023).
%GWs
%Scalar induced gravitational waves in modified teleparallel gravity theories

\bibitem{Braglia:2021NEWBraglia}
M. Braglia, X. Chen, and D. K. Hazra, J. Cosmol. Astropart. Phys. {\bf 03}, 005 (2021).
%SGWB
%Probing primordial features with the stochastic gravitational wave background

\bibitem{Braglia:2022NEWBraglia}
M. Braglia, et al., J. Cosmol. Astropart. Phys. {\bf 08}, 001 (2020).
%SGWB
%Generating PBHs and small-scale GWs in two-field models of inflation
%Matteo Braglia1,2,3, Dhiraj Kumar Hazra2,3,4, Fabio Finelli2,3, George F. Smoot5,6,7,8, L. Sriramkumar9 and Alexei A. Starobinsky10

\bibitem{Braglia:2024NEWBraglia}
M. Braglia, et al., arXiv:2407.04356.
%LISA
%Gravitational waves from inflation in LISA: reconstruction pipeline and physics interpretation
%The impact of a primordial gravitational wave background on LISA resolvable sources%Matteo Braglia, Mauro Pieroni, Sylvain Marsat %arXiv:2406.10048









%GWs
\bibitem{BartoloPRL:2019}
N. Bartolo et al., Phys. Rev. Lett. {\bf 122}, 211301 (2019).
%
\bibitem{BartoloPRD:2019}
N. Bartolo et al., Phys. Rev. D {\bf 99}, 103521 (2019).
%
\bibitem{CaiPRL:2019}
R. G. Cai, S. Pi, and M. Sasaki, Phys. Rev. Lett. {\bf 122}, 201101 (2019).
%
\bibitem{Fumagalli:2021}
J. Fumagalli, S. Renaux-Petel, and L. T. Witkowski, J. Cosmol. Astropart. Phys. {\bf 08}, 030
(2021).
%
\bibitem{CaiJCAP:2019}
R. G. Cai, S. Pi, S. J. Wang, and X. Y. Yang, J. Cosmol. Astropart. Phys. {\bf 05}, 013 (2019).
%
\bibitem{CaiPRD:2019}
Y. F. Cai et al., Phys. Rev. D {\bf 100}, 043518 (2019).
%
\bibitem{Wang:2019}
S. Wang, T. Terada, and K. Kohri, Phys. Rev. D {\bf 99}, 103531 (2019).
%

%NanoGrav
\bibitem{Agazie1:2023}
G. Agazie et al., (NANOGrav Collaboration), Astrophys. J. Lett. {\bf 951}, L8 (2023).
%
\bibitem{Agazie2:2023}
G. Agazie et al., (NANOGrav Collaboration), Astrophys. J. Lett. {\bf 951}, L9 (2023).
%
\bibitem{Agazie3:2023}
G. Agazie et al., (NANOGrav Collaboration), Astrophys. J. Lett. {\bf 951}, L10 (2023).
%
\bibitem{Agazie4:2023}
G. Agazie et al., (NANOGrav Collaboration), Astrophys. J. Lett. {\bf 951}, L11 (2023).
%
%
\bibitem{Ragavendra:2023}
H. V. Ragavendra et al., Galaxies {\bf 11}, 34 (2023).
%
\bibitem{Dimopoulos:2017}
K. Dimopoulos, Phys. Lett. B {\bf775}, 262 (2017).
%
\bibitem{Lanczos:1938}
C. Lanczos, Ann. Math. {\bf 39}, 842 (1938).
%Annals of Mathematics, Second Series {\bf 39}, 4, 842-850 (1938).
%15No	
%
\bibitem{Lovelock:1971}
D. Lovelock, and J. Math. Phys. {\bf 12}, 498 (1971).
%16No
%
\bibitem{RashidiNozari:2020}
N. Rashidi, and K. Nozari, Astrophys. J. {\bf 890}, 58 (2020).
%
\bibitem{AziziNozari:2022}
S. Azizi, S. Eslamzadeh, J. T. Firouzjaee, and K. Nozari, Nucl. Phys. B {\bf 985}, 115993 (2022).
%
\bibitem{ShahraeiniNozari:2022}
S. S. Shahraeini, K. Nozari, and S. Saghafi, J. Hologr. Appl. Phys. {\bf 2}, 55 (2022).
%Journal of Holography Applications in Physics
%
\bibitem{Myers:1988}
R. C. Myers, and J. Z. Simon, Phys. Rev. D {\bf 38}, 2434 (1988).
%Robert C. Myers, Jonathan Z. Simon, Phys. Rev. D {\bf 38}, 2434 (1988).
%18No

\bibitem{Boulware:1985}
D. G. Boulware, and S. Deser, Phys. Rev. Lett. {\bf 55}, 2656 (1985).
%David G. Boulware, S. Deser, Phys. Rev. Lett. {\bf 55}, 2656 (1985).
%17No

\bibitem{Sahabandu:2006}
C. Sahabandu, P. Suranyi, C. Vaz, and L. C. R. Wijewardhana, Phys. Rev. D {\bf 73}, 044009 (2006).
%C. Sahabandu, P. Suranyi, C. Vaz, and L. C. R. Wijewardhana, Phys. Rev. D {\bf 73}, 044009 (2006).
%20No	

\bibitem{Cho:2002}
Y. M. Cho, and I. P. Neupane, Phys. Rev. D {\bf 66}, 024044 (2002).
%Y. M. Cho, and Ishwaree P. Neupane, Physical Review D {\bf 66}, 024044(2002).
%19No

\bibitem{Ghosh:2014}
S. G. Ghosh, S. D. Maharaj, Phys. Rev. D {\bf 89}, 084027 (2014).	
%Sushant G. Ghosh, Sunil D. Maharaj, Phys. Rev. D {\bf 89}, 084027 (2014).
%22No

\bibitem{HabibMazharimousavi:2009}
S. H. Mazharimousavi, and M. Halilsoy, Phys. Lett. B {\bf 681}, 190 (2009).
%S. Habib Mazharimousavi and M. Halilsoy, Phys. Lett. B {\bf 681}, 190 (2009).
%21No
%



\bibitem{Das:2019}
S. Das, Phys. Rev. D \textbf{99}, 083510 (2019).
%
\bibitem{Garg:2019}
{\bf S. K. Garg, and C. Krishnan, J. High Energy Phys. {\bf11}, 075 (2019).}
%
\bibitem{Ooguri:2019}
H. Ooguri, E. Palti, G. Shiu, and C. Vafa, Phys. Lett. B {\bf788}, 180 (2019).
%
\bibitem{Kehagias:2018}
A. Kehagias, and A. Riotto, Fortschr. Phys. {\bf66}, 1800052 (2018).







\bibitem{EOPozdeeva:2020}
E. O. Pozdeeva, M. R. Gangopadhyay, M. Sami et al., Phys. Rev. D {\bf 102}, 043525 (2020).
%Pozdeeva EO, Gangopadhyay MR, Sami M, Toporensky AV, Vernov SY. Inflation with a quartic potential in the framework of Einstein-Gauss-Bonnet gravity. Physical Review D. 2020 Aug 28;102(4):043525.     GB GOOOOOOOOOOOD
		
\bibitem{HAKhan:2022}
H. A. Khan, Phys. Rev. D {\bf 105}, 063526 (2022).
%H. A. Khan, Study of Goldstone inflation in the domain of Einstein-Gauss-Bonnet gravity. Physical Review D. 2022 Mar 24;105(6):063526.     GB GOOOOOOOOOOOD
		
\bibitem{MRGangopadhyay:2023a}
M. R. Gangopadhyay, and H. A. Khan, Phys. Dark Universe {\bf 40}, 101177 (2023).
%Gangopadhyay MR, Khan HA. A case study of small field inflationary dynamics in the Einstein–Gauss–Bonnet framework in the light of GW170817. Physics of the Dark Universe. 2023 May 1;40:101177.     GB GOOOOOOOOOOOD
		
\bibitem{SDOdintsov:2020}
S. D. Odintsov, V. K. Oikonomou, and F. P. Fronimos, Ann. Phys. {\bf 420}, 168250 (2020).
%S.D. Odintsov, V. K. Oikonomou and F. P. Fronimos,Non-minimally coupled Einstein–Gauss–Bonnet inflation phenomenology in view of GW170817. Annals of Physics. 2020 Sep 1;420:168250.     GB
		
\bibitem{SDOdintsov:2023}
S. D. Odintsov, V. K. Oikonomou, and F. P. Fronimos, Phys. Rev. D {\bf 107}, 084007 (2023).
%Odintsov SD, Oikonomou VK, Fronimos FP. Inflationary dynamics and swampland criteria for modified Gauss-Bonnet gravity compatible with GW170817. Physical Review D. 2023 Apr 4;107(8):084007.      GB
		
		
\bibitem{SNojiri:2023a}
S. Nojiri, S. D. Odintsov, and D. S. G{\`o}mez, Phys. Dark Universe {\bf 41}, 101238 (2023).
%Nojiri SI, Odintsov SD, Gómez DS. Unifying inflation with early and late dark energy in Einstein–Gauss–Bonnet gravity. Physics of the Dark Universe. 2023 Aug 1;41:101238.      GB
%
%
%FPBH FIGURE
\bibitem{Jiang:2013}
P. X. Jiang, J. W. Hu, and Z. K. Guo, Phys. Rev. D {\bf 88}, 123508 (2013).
%
\bibitem{Koh:2014}
S. Koh, Phys. Rev. D {\bf 90}, 063527 (2014).
%
\bibitem{Guo:2010}
Z. K. Guo, Phys. Rev. D {\bf 81}, 123520 (2010).
%	
\bibitem{Odintsov:2020}
S. D. Odintsov, V. K. Oikonomou, and F. P. Fronimos, Nucl. Phys. B {\bf 958}, 115135 (2020).
%
\bibitem{Gao:2020}
T. J. Gao, Eur. Phys. J. C {\bf 80}, 1013 (2020).
%
\bibitem{Felice:2011}
A. De. Felice, and S. Tsujikawa, J. Cosmol. Astropart. Phys. {\bf 04}, 029 (2011).		
%
\bibitem{Odintsov:2020nsr}
S. D. Odintsov, and V. K. Oikonomou, Phys. Lett. B {\bf 805}, 135437 (2020).
%		
\bibitem{Ps&PtHorndeski}
G. W. Horndeski, Int. J. Theor. Phys. {\bf 10}, 363 (1974).
%
\bibitem{Campeti:2022}
P. Campeti, and E. Komatsu, Astrophys. J. {\bf 941}, 110 (2022).
%
\bibitem{Ade:2021}	
P. A. R. Ade et al., (BICEP/Keck Collaboration), Phys. Rev. Lett. {\bf 127}, 151301 (2021).
%

\bibitem{Dvali:1994}
G. R. Dvali, Q, Shafi, and R. K. Schaefer, Phys. Rev. Lett., {\bf 73}, 1886 (1994).
%
\bibitem{Copeland:1994}
E. J. Copeland, A. R. Liddle, D. H. Lyth, E. D. Stewart, and D. Wands, Phys. Rev. D, {\bf 49}, 6410 (1994).
%
\bibitem{Linde:1994}
A. D. Linde, Phys. Rev. D, {\bf 49}, 748 (1994).
%
\bibitem{Fixsen:1996}
D. J. Fixsen et al., Astrophys. J. {\bf 473}, 576 (1996).
%
\bibitem{Inomata:2016}
K. Inomata, M. Kawasaki, and Y. Tada, Phys. Rev. D {\bf 94}, 043527 (2016).
%
\bibitem{Inomata:2019-a}
K. Inomata, and T. Nakama, Phys. Rev. D {\bf 99}, 043511 (2019).	
%
\bibitem{Tada:2019}
Y. Tada, and  S. Yokoyama, Phys. Rev. D {\bf100}, 023537 (2019).
%77
%
\bibitem{young:2014}
S. Young, C. T. Byrnes, and M. Sasaki, J. Cosmol. Astropart. Phys. {\bf 07}, 045 (2014).
%
\bibitem{Harada:2013}
T. Harada, C. M. Yoo, and K. Kohri, Phys. Rev. D {\bf88}, 084051 (2013).
%80
%
\bibitem{Musco:2013}
I. Musco, and J. C. Miller, Class. Quant. Grav. {\bf 30}, 145009 (2013).
%

\bibitem{CMB}
P. D. Serpico, V. Poulin, D. Inman, and K. Kohri,  Phys. Rev. Res. {\bf2}, 023204 (2020).
%
\bibitem{Kavanagh:2018}
B. J. Kavanagh, D. Gaggero, and G. Bertone, Phys. Rev. D {\bf 98}, 023536 (2018).

\bibitem{Chen:2022}
Z. C. Chen, Y. M. Wu, and Q. G. Huang, Astrophys. J. {\bf 936}, 20 (2022).		
\bibitem{Boehm:2021}
C. Boehm et al., J. Cosmol. Astropart. Phys. {\bf 03}, 078 (2021).
%81

\bibitem{EORS}
P. Tisserand et al., Astron. Astrophys. {\bf 469}, 387 (2007).
%
\bibitem{Kepler}
K. Griest, A. M. Cieplak, and  M. J. Lehner, Astrophys. J. {\bf 786}, 158 (2014).
%
\bibitem{Icarus}
M. Oguri et al., Phys. Rev. D {\bf 97}, 023518 (2018).
%
\bibitem{subaro}
D. Croon, D. McKeen, N. Raj, and Z. Wang, Phys. Rev. D {\bf 102}, 083021 (2020).
%87
\bibitem{EGG}
B. J. Carr et al., Phys. Rev. D {\bf 81}, 104019 (2010).
%
\bibitem{Laha:2019}
R. Laha, Phys. Rev. Lett. {\bf123}, 251101 (2019).
%		
\bibitem{Clark}
S. Clark et al., Phys. Rev. D {\bf 95}, 083006 (2017).
%	
%
%GWs
\bibitem{Matarrese:1998}
S. Matarrese, S. Mollerach, and M. Bruni, Phys. Rev. D {\bf 58}, 043504 (1998).
%
\bibitem{Mollerach:2004}
S. Mollerach, D. Harari, and S. Matarrese, Phys. Rev. D {\bf 69}, 063002 (2004).
%
\bibitem{Saito:2009}
R. Saito, and J. Yokoyama, Phys. Rev. Lett. {\bf 102}, 161101 (2009.)
%
\bibitem{Garcia:2017}
J. Garcia-Bellido, M. Peloso, and C. Unal, J. Cosmol. Astropart. Phys. {\bf 09}, 013 (2017).
%
\bibitem{Kohri:2018}
K. Kohri, and T. Terada, Phys. Rev. D {\bf 97}, 123532 (2018).
%
\bibitem{Cai:2019-a}
R. G. Cai, S. Pi, and M. Sasaki, Phys. Rev. Lett. {\bf 122}, 201101 (2019).
%
\bibitem{Cai:2019-b}
R. G. Cai, S. Pi, S. J. Wang, and X. Y. Yang, J. Cosmol. Astropart. Phys. {\bf 05}, 013 (2019).
%
\bibitem{Cai:2019-c}
Y. F. Cai et al., Phys. Rev. D {\bf 100}, 043518 (2019).
%
\bibitem{Bartolo:2019-a}
N. Bartolo et al., Phys. Rev. Lett. {\bf 122}, 211301 (2019).
%
\bibitem{Bartolo:2019-b}
N. Bartolo et al., Phys. Rev. D {\bf 99}, 103521 (2019).
%
\bibitem{Fumagalli:2020b}
J. Fumagalli, S. Renaux-Petel, and L. T. Witkowski, J. Cosmol. Astropart. Phys. {\bf 08}, 030  (2021).
%
\bibitem{Hajkarim:2019}
F. Hajkarim, and J. Schaffner-Bielich, Phys. Rev. D {\bf 101}, 043522 (2020).
%
\bibitem{Xu:2020}
W. T. Xu, J. Liu, T. J. Gao, and Z. K. Guo, Phys. Rev. D {\bf 101}, 023505 (2020).
%
\bibitem{Fu:2020}
C. Fu, P. Wu, and H. Yu, Phys. Rev. D {\bf 101}, 023529 (2020).
%
\bibitem{Domenech:2020a}
G. Dom{\`e}nech, and M. Sasaki, Int. J. Mod. Phys. D {\bf 29}, 2050028 (2020).
%
\bibitem{Domenech:2020b}
G. Dom{\`e}nech, and M. Sasaki, Phys. Rev. D {\bf 103}, 063531 (2021).
%
%
\bibitem{Lu:2019}
Y. Lu, Y. Gong, Z. Yi, and F. Zhang, J. Cosmol. Astropart. Phys. {\bf 12}, 031 (2019).
%
\bibitem{Ananda:2007}
K. N. Ananda, C. Clarkson, and D. Wands, Phys. Rev. D {\bf 75}, 123518 (2007).
%
\bibitem{Baumann:2007}
D. Baumann et al., Phys. Rev. D {\bf 76}, 084019 (2007).
%
%
%GWs FIGURE
\bibitem{ligo-a}
G. M. Harry (LIGO Scientific Collaboration), Class. Quant. Grav. {\bf 27}, 084006 (2010).
%
\bibitem{ligo-b}
J. Aasi et al., (LIGO Scientific Collaboration), Class. Quant. Grav. {\bf 3}2, 074001 (2015).
%
\bibitem{lisa-a}
K. Danzmann, Class. Quant. Grav. {\bf 14}, 1399 (1997).
%
\bibitem{lisa}
P. Amaro-Seoane et al., (LISA Collaboration), arXiv:1702.00786.
%
\bibitem{Harry:2006BBO}
G. M. Harry et al., Class. Quant. Grav. {\bf 23}, 4887 (2006).
%	
\bibitem{Crowder:2005BBO}
J. Crowder, and N. J. Cornish, Phys. Rev. D {\bf 72}, 083005 (2005).
%	
\bibitem{Corbin:2006BBO}
V. Corbin, and N. J. Cornish, Class. Quant. Grav. {\bf 23}, 2435 (2006).
%	
\bibitem{Yagi:2011BBODECIGO}
K. Yagi, and N. Seto, Phys. Rev. D {\bf 83}, 044011 (2011).
%	
\bibitem{Yagi:2017BBODECIGO}
K. Yagi, and N. Seto, Phys. Rev. D {\bf 95}, 109901 (2017).
%
\bibitem{Kawamura:2006DECIGO}
S. Kawamura et al., Class. Quant. Grav. {\bf 23}, S125 (2006).
%		
\bibitem{Kawamura:2011DECIGO}
S. Kawamura et al., Class. Quant. Grav. {\bf 28}, 094011 (2011).
%
\bibitem{Seto:2001DECIGO}
N. Seto, S. Kawamura, and T. Nakamura, Phys. Rev. Lett. {\bf 87}, 221103 (2001).	
%
\bibitem{ska}
C. J. Moore, R. H. Cole, and C. P. L. Berry, Class. Quant. Grav. {\bf 32}, 015014 (2015).
%
\bibitem{skaCarilli:2004}
C. L. Carilli, and S. Rawlings, New Astron. Rev. {\bf 48}, 979 (2004).
%
\bibitem{skaWeltman:2020}
A. Weltman et al., Publ. Astron. Soc. Aust. {\bf 37}, e002 (2020).	
%
%
%EPTA
\bibitem{EPTA-a}
R. D. Ferdman et al., Class. Quant. Grav. {\bf 27}, 084014 (2010).
%
\bibitem{EPTA-b}
G. Hobbs et al., Class. Quant. Grav. {\bf 27}, 084013 (2010).
%
\bibitem{EPTA-c}
M. A. McLaughlin, Class. Quant. Grav. {\bf 30}, 224008 (2013).
%
\bibitem{EPTA-d}
G. Hobbs, Class. Quant. Grav. {\bf 30}, 224007 (2013).
%
%
%
%GWs Fit f
\bibitem{Fu:2019vqc}
C. Fu, P. Wu, and H. Yu, Phys. Rev. D {\bf 101}, 023529 (2020).
%		
\bibitem{Fu:2020lob}
C. Fu, P. Wu, and H. Yu, Phys. Rev. D {\bf 102}, 043527 (2020).
%
\bibitem{Kuroyanagi}
S. Kuroyanagi, T. Chiba, and T. Takahashi, J. Cosmol. Astropart. Phys. {\bf 11}, 038 (2018).
%		
\bibitem{Xu}
W. T. Xu, J. Liu, T. J. Gao, and Z. K. Guo, Phys. Rev. D {\bf 101}, 023505 (2020).
%
\bibitem{Bagui:2021dqi}
E. Bagui, and S. Clesse, Phys. Dark Universe {\bf 38}, 101115 (2022).
%		
\bibitem{Sasaki:2020}
R. G. Cai, S. Pi, and M. Sasaki, Phys. Rev. D {\bf 102}, 083528 (2020).
%			
\bibitem{Yuan:2020}
C. Yuan, Z. C. Chen, and Q. G. Huang, Phys. Rev. D {\bf 101}, 043019 (2020).
%
%
%NanoGrav Fit
\bibitem{VagnozziFitNANOGrav:2023}
S. Vagnozzi, J. High Energy Astrophys. {\bf 39}, 81 (2023).
%
\bibitem{BabichevFitNANOGrav:2023}
E. Babichev, D. Gorbunov, S. Ramazanov, R. Samanta, and A. Vikman, Phys. Rev. D {\bf 108}, 123529 (2023).
%E. Babichev, D. Gorbunov, S. Ramazanov, R. Samanta, and A. Vikman Phys. Rev. D 108, 123529 – Published 18 December 2023  arXiv:2307.04582
%
\bibitem{AntoniadisFitNANOGrav:2023a}
J. Antoniadis, P. Arumugam, S. Arumugam, et al., Astron. Astrophys. {\bf 678}, A48 (2023).
%
\bibitem{AntoniadisFitNANOGrav:2023b}
J. Antoniadis, P. Arumugam, S. Arumugam, et al., Astron. Astrophys. {\bf 678}, A50 (2023).
%
\bibitem{AntoniadisFitNANOGrav:2023c}
J. Antoniadis, P. Arumugam, S. Arumugam, et al., Astron. Astrophys. {\bf 685}, A94 (2024).
%Astronomy & Astrophysics 685 (2024): A94.		arXiv:2306.16227
%

	\end{thebibliography}
\end{document}